\documentclass[acmsmall]{acmart}

\acmJournal{PACMMOD}
\acmYear{2025} \acmVolume{3} \acmNumber{6 (SIGMOD)}
\acmArticle{XXX} \acmMonth{12} \acmPrice{}
\acmDOI{10.1145/XXXXXXX}

\usepackage{pifont}
\usepackage{graphicx}
\usepackage{booktabs}
\usepackage{multirow}
\usepackage{enumitem}
\usepackage{listings}
\usepackage[ruled,vlined]{algorithm2e}

\usepackage{tabularx} 
\usepackage{subfig}

\theoremstyle{acmdefinition}
\newtheorem{definition}{Definition}[section]
\newtheorem{example}{Example}[section]

\begin{document}

\title{SQLBarber: A System Leveraging Large Language Models to Generate Customized and Realistic SQL Workloads}

\lstset{
  language       = SQL,
  basicstyle     = \ttfamily\small,
  numbers        = left,
  numberstyle    = \small,
  breaklines     = true,
  frame          = single,
  showstringspaces = false,
  xleftmargin    = 1em,
  keywordstyle   = \ttfamily\bfseries,
  mathescape     = true,               
  moredelim=**[is][\color{blue}]{~}{~},  
  moredelim=**[is][\color{red}]{-}{-},   
}

\author{Jiale Lao}
\affiliation{%
  \institution{Cornell University}
  \city{Ithaca, New York}
  \country{USA}}
\email{jiale@cs.cornell.edu}

\author{Immanuel Trummer}
\affiliation{%
  \institution{Cornell University}
  \city{Ithaca, New York}
  \country{USA}}
\email{itrummer@cornell.edu}


\begin{abstract}
Database research and development often require a large number of SQL queries for benchmarking purposes. However, acquiring real-world SQL queries is challenging due to privacy concerns, and existing generation methods offer limited options for customization and for satisfying realistic constraints.
To address this issue, we present SQLBarber, a system based on Large Language Models (LLMs) to generate customized and realistic SQL workloads. SQLBarber (1) eliminates the need for users to manually craft SQL templates in advance, while providing the flexibility to accept natural language specifications to constrain SQL templates, (2) scales efficiently to generate large volumes of queries matching any user-defined cost distribution (e.g., cardinality and execution plan cost), and (3) uses execution statistics from production environments 
to extract SQL template specifications and query cost distributions that reflect real-world query characteristics. SQLBarber introduces (1) a declarative interface for users to effortlessly generate customized SQL templates, (2) an LLM-powered pipeline augmented with a self-correction module that profiles, refines, and prunes SQL templates based on query costs, and (3) a Bayesian Optimizer to efficiently explore predicate values and identify a set of queries that satisfy the target cost distribution. We construct and open-source ten benchmarks of varying difficulty levels and target query cost distributions based on real-world statistics from Snowflake and Amazon Redshift. Extensive experiments on these benchmarks show that SQLBarber is the only system that can generate customized SQL templates. It reduces query generation time by one to two orders of magnitude and significantly improves alignment with the target cost distribution, compared with existing methods.

\end{abstract}

\maketitle

\section{Introduction}


\textcolor{black}{Leading companies such as Amazon and Snowflake have recently released workload statistics to facilitate the construction of realistic benchmarks~\cite{redset-vldb, snowflake-nsdi20}. However, due to privacy constraints, the actual queries and underlying databases are unavailable; only high-level workload characteristics are shared. These typically include: (1) properties of SQL templates (e.g., the number of accessed tables and the distributions of query operators such as joins and aggregations), and (2) cost distributions of instantiated queries (e.g., the distributions of cardinalities, execution times, and CPU usages). This raises an important research direction: transforming high-level workload information into realistic and executable benchmarks.}

\textcolor{black}{Recent methods have been proposed for generating SQL queries, but they struggle to produce customized queries that satisfy realistic constraints. SQLSmith~\cite{sqlsmith}, SQLancer~\cite{ba2023testing}, and SQLStorm~\cite{sqlstorm} focus on generating diverse queries for testing and debugging. However, these methods do not allow users to control SQL templates or query costs, since their generation processes rely on randomization to ensure diversity. HillClimbing~\cite{hillclimb, cost-aware-query-generation} and LearnedSQLGen~\cite{learnedsqlgen} support constraints on query costs but not on SQL templates, and their cost constraints are not derived from real-world statistics. CAB~\cite{CAB} and RedBench~\cite{krid2025redbench} synthesize workloads based on real-world statistics~\cite{redset-vldb, snowflake-nsdi20}. However, they do not generate new queries; instead, they reuse queries from existing benchmarks, which have been shown to differ from real-world workloads~\cite{redset-vldb, 10.1145/3209950.3209952}.} 



Thus we present SQLBarber, a system based on Large Language Models (LLMs) to generate customized and realistic SQL workloads. SQLBarber (1) removes the need for users to manually create SQL templates in advance, while allowing them to specify high-level natural language constraints for customized SQL template generation (e.g., number of tables, joins, aggregations, or presence of certain operators and complex scalar expressions), (2) scales efficiently to produce a large number of queries that match any user-defined cost distribution (e.g., cardinality and execution plan cost), and (3) uses execution statistics collected from Amazon Redshift \cite{redset-vldb} and Snowflake \cite{snowflake-nsdi20} to derive both the specifications for SQL templates and the cost distributions of SQL queries, ensuring that the generated SQL workloads reflect production environments.

\color{black}
\begin{example}
A database researcher is developing a novel caching method and wants to benchmark it on realistic workloads from industry. Having high-level information about workload properties, including information on query template structure as well as query cost distribution, as released by companies such as Snowflake, the researcher can generate one or multiple sets of templates with associated queries using SQLBarber. SQLBarber ensures that the generated workloads are close to the target workloads, meaning that benchmark results obtained for the novel caching method and baselines are meaningful.
\end{example}
\color{black}

SQLBarber employs an LLM-powered SQL Template Generator to create customized SQL templates for a target database based on user-defined natural language specifications. It connects to the target database to generate a textual summary of the database schema and identify possible join paths. The generator then constructs a prompt to combine the above information and invoke LLMs to generate SQL templates. Due to the hallucination problem of LLMs (i.e., LLMs can generate plausible but factually incorrect or nonsensical information~\cite{xu2025hallucinationinevitableinnatelimitation, 10.1145/3703155}), the generated SQL templates could have syntax errors and do not satisfy user specifications. Therefore, we propose using the reasoning and self-correction ability of LLMs to make this generation process reliable. Specifically, the generator uses LLMs to judge whether an SQL template satisfies the constraints and provide the reasoning process. If the template fails, the generator invokes LLMs to iteratively rewrite this template based on the reasoning results and error messages from the DBMS. Finally, the generator ensures that \textcolor{black}{the generated SQL templates are executable after instantiation} and compliant with user specifications.

SQLBarber relies on a Cost-Aware Query Generator to efficiently produce a large number of queries constrained to a target cost distribution. It takes as input the generated SQL templates and any user-specified cost distribution (e.g., cardinality, execution plan cost, or execution time). Firstly, it employs a profiling stage to evaluate the capability of each SQL template to produce queries of different costs. This is achieved by instantiating the previously generated SQL templates with different predicate values and evaluating these queries on the DBMS. Secondly, based on the profiling statistics, it refines these seed SQL templates to create new templates that can fill the cost ranges that are not covered by existing templates, and prunes templates that cannot contribute to the target cost distribution. Finally, it utilizes Bayesian Optimization (BO)~\cite{JMLR:v23:21-0888} to efficiently explore the predicate value space and identify a set of SQL queries that satisfy the target cost distribution.

SQLBarber leverages real-world execution statistics to ensure that the generated SQL workloads accurately reflect production environments. To the best of our knowledge, it is the first system to do so. Specifically, SQLBarber analyzes the statistics collected from Amazon Redshift \cite{redset-vldb} and Snowflake \cite{snowflake-nsdi20} to derive both the specifications for SQL templates and the cost distributions of SQL queries. The specifications define different features for different SQL templates, including the number of tables to access, the number of joins and aggregations, and the presence of certain operators. These specifications are combined with user-provided customized requirements to constrain the SQL template generation process of SQLBarber. Moreover, SQLBarber can generate SQL queries whose cost distribution closely matches the cost distribution observed in real-world workloads.
We derive cardinality and execution time distributions from the query logs published by Amazon Redshift and Snowflake, and use these distributions to control the query generation process of SQLBarber. It is important to note that SQLBarber is not restricted to these specific distributions, and can generate queries that follow any user-specified cost distribution.

We extensively evaluate SQLBarber and compare it to state-of-the-art SQL generation methods. We construct and open-source ten realistic benchmarks for SQL workload generation, where each benchmark is characterized by the constraints on SQL templates and the cost distributions of SQL queries. 
All constraints are extracted from real-world statistics published by Amazon Redshift and Snowflake. We use TPC-H and IMDB as the underlying databases and deploy PostgreSQL as the query execution engine.
We classify these benchmarks into different difficulties based on the dataset source, the number of queries to generate, and the number of intervals to split the target cost range. Extensive experiments on these benchmarks show that SQLBarber achieves one to two orders of magnitude improvement in query generation time, and shows significantly better alignment between the target cost distribution and the cost distribution of generated queries, compared to existing methods~\cite{learnedsqlgen, hillclimb}. Moreover, SQLBarber is the only system that can create customized SQL templates based on natural language instructions by users. We also conduct thorough scalability, ablation, and cost studies to evaluate different aspects of SQLBarber. The code is available at \textcolor{blue}{\url{https://github.com/SolidLao/SQLBarber}} and a demonstration is available at ~\cite{sqlbarber-demo}. 

Our original scientific contributions are the following:
\begin{itemize}
    \item We present SQLBarber, a system based on LLMs to generate customized and realistic SQL workloads.
    \item We develop a reliable SQL template generator with a self-correction mechanism to create customized SQL templates based on natural language instructions and DBMS feedback.
    \item We propose a cost-aware query generator that efficiently produces a large number of queries to fill any cost distribution by profiling, refining, pruning, and instantiating SQL templates under the guidance of Bayesian Optimization. 
    \item We construct and release ten realistic benchmarks for SQL workload generation by analyzing real-world execution statistics collected from Amazon Redshift and Snowflake. 
    \item We evaluate SQLBarber experimentally on the proposed benchmarks and compare against state-of-the-art baselines.
\end{itemize}

The remainder of this paper is organized as follows.
\textcolor{black}{We discuss related work in Section~\ref{subsec: related-work}.} We formally define the SQL workload generation problem in Section~\ref{sec:problem-formulation} and give an overview of SQLBarber in Section~\ref{sec:overview}. Section~\ref{sec:template-generator} describes the customized SQL template generator, and Section~\ref{sec:query-generator} discusses the cost-aware query generator. We conduct extensive experiments to evaluate the performance, scalability, and robustness of SQLBarber with an ablation study, a cost analysis, \textcolor{black}{and extensive further analysis} in Section~\ref{sec:experi}. We conclude in Section~\ref{sec:conclusion}.

\section{\textcolor{black}{Related Work}}\label{subsec: related-work}

\noindent \textbf{\textcolor{black}{Diversity-Oriented Query Generation.}} \textcolor{black}{SQLBarber relates to prior work on generating diverse SQL queries for testing and debugging~\cite{sqlsmith, sqlstorm, ba2023testing}.
SQLSmith~\cite{sqlsmith} recursively and randomly expands grammar rules while ensuring schema and type consistency.
SQLancer~\cite{ba2023testing} employs a feedback-guided test generation approach that aims to produce diverse query execution plans, based on the idea that exploring a wide range of plans helps reveal bugs.
SQLStorm~\cite{sqlstorm} directly prompts LLMs to generate SQL queries with a high temperature to increase diversity.
While the aforementioned methods emphasize query diversity for testing without constraints on SQL templates or query costs, SQLBarber integrates constraints on both templates and costs to turn high-level workload information into realistic, executable benchmarks.}


\noindent \textbf{\textcolor{black}{Constraint-Aware Query Generation.}} \textcolor{black}{SQLBarber is also related to prior work on cost-constrained SQL query generation~\cite{hillclimb, cost-aware-query-generation, learnedsqlgen}. Prior work~\cite{hillclimb, cost-aware-query-generation} uses the hill climbing algorithm to instantiate given SQL templates by adjusting predicate values so that the resulting queries meet target cost ranges (e.g., cardinalities within $[1000, 2000]$). LearnedSQLGen~\cite{learnedsqlgen} extends this idea by applying reinforcement learning (RL) to explore both SQL templates and predicate values. However, none of these approaches generate new templates according to user specifications. In contrast, SQLBarber can generate new and customized SQL templates guided by real-world workload statistics~\cite{redset-vldb, snowflake-nsdi20}.} 

\noindent \textbf{\textcolor{black}{Analysis of Industry Workloads.}}
\textcolor{black}{
SQLBarber builds on prior work that analyzes industrial workloads~\cite{redset-vldb, snowflake-nsdi20, alibaba-template-analysis}. Redset~\cite{redset-vldb} and Snowset~\cite{snowflake-nsdi20} present detailed statistics on production workloads from Amazon Redshift and Snowflake, respectively. These datasets provide statistics for both SQL templates and query costs. Prior studies further show that query repetition is a key characteristic of real-world workloads~\cite{redset-vldb, alibaba-template-analysis}. For example, Amazon Redshift reports that approximately 85\%–90\% of executed queries originate from repeated templates in 50\% of its clusters~\cite{redset-vldb}, while Alibaba’s logs show that 4.5 million queries correspond to only 205 templates~\cite{alibaba-template-analysis}. Motivated by these findings, SQLBarber adopts templated query generation by first creating new templates from scratch using LLMs and then instantiating queries through Bayesian Optimization (BO). Both steps are guided by real-world statistics~\cite{redset-vldb, snowflake-nsdi20} to ensure that the generated workloads remain realistic.}

\noindent \textbf{\textcolor{black}{Real-World Statistics-Guided Workload Synthesis.}} \textcolor{black}{SQLBarber relates to prior works that leverage real-world statistics (Redset~\cite{redset-vldb} and Snowset~\cite{snowflake-nsdi20}) to construct SQL workloads.
CAB~\cite{CAB} uses the original 22 TPC-H templates (with insert and delete streams), deploys multiple databases, and schedules query arrivals based on a real-world Snowflake trace~\cite{snowflake-nsdi20} to model multi-tenancy and workload fluctuations in cloud databases.
RedBench~\cite{krid2025redbench} samples queries from existing benchmarks according to the number of normalized joins while preserving the query repetition patterns observed in Amazon Redshift workloads~\cite{redset-vldb}.
However, these approaches do not generate new queries; they reuse or sample existing templates and are thus limited by the diversity and structure of available benchmarks, which differ from real-world workloads~\cite{redset-vldb, 10.1145/3209950.3209952}.
They primarily reproduce execution statistics with limited control over query templates—CAB does not model template structures, and RedBench only considers the number of normalized joins.
In contrast, SQLBarber generates new and customizable SQL queries, constraining both template structures and query costs using real-world workload statistics~\cite{redset-vldb, snowflake-nsdi20}, ensuring both realism and flexibility.} 

\noindent \textbf{LLM-Enhanced DBMS.} More broadly, SQLBarber relates to prior works on leveraging LLMs to enhance DBMS, including DBMS tuning \cite{10.14778/3659437.3659449, 10.1145/3626246.3654751, 10.1145/3733620.3733641, gptunerdemo, huang2025e2etuneendtoendknobtuning, li2024largelanguagemodelgood, 10.1145/3514221.3517843}, query rewrite~\cite{10.14778/3551793.3551841, song2025quitequeryrewriterules, 10.14778/3696435.3696440}, natural language interface \cite{10.1145/3654930, lin2025toxicsqlmigratingsqlinjection, 10.1145/3626246.3654732}, and semantic query processing~\cite{lao2025sembenchbenchmarksemanticquery, thalamusdb, liu2025palimpzest}.


\section{\textcolor{black}{Problem Formulation}} \label{sec:problem-formulation}

\textcolor{black}{Since the actual queries and underlying databases are inaccessible due to privacy constraints, the goal of the customized and realistic SQL workload generation problem is to transform high-level workload information into an executable workload whose SQL template properties and query cost distributions closely match those of the original production workload. In this process, both template generation and query instantiation are guided by production workload statistics. Next, we formally define the problem.}

\begin{definition}[Database]
\label{definition:database}
\textcolor{black}{The user must specify a target database $\mathbf{D}=\{R_1,R_2,\dots,R_n\}$ on which to generate queries. Depending on privacy requirements, this can be a production database with a public schema and masked data, or a database that differs from the production database from which constraints are derived. The SQL generation method should produce SQL templates and instantiate queries whose workload statistics match the target as closely as possible, even when operating on a different database.} \textcolor{black}{Following prior work on query or benchmark generation~\cite{sqlstorm, learnedsqlgen, hillclimb, CAB, krid2025redbench}, this problem does not constrain the underlying database, and existing methods typically use one from available benchmarks.} \textcolor{black}{Although we adopt the same assumption, our experiments in Section~\ref{subsec: further-anlaysis} show that (1) SQLBarber can adjust its generation process to adapt to different database types and sizes, and (2) the selected database affects the diversity and structure of the generated queries. This finding suggests an interesting direction for future work: integrating learned database generation~\cite{database-generation, db-generation, SAM-data-generation} with query generation to enable end-to-end benchmark customization that jointly models both databases and queries.}
\end{definition}



\begin{definition}[SQL Template]
\label{definition: customized-templates}
SQL templates $\mathbf{T} = \{t_1, t_2, \dots, t_n\}$ are predefined SQL statements in which certain components of each template $t_i$ are placeholders $\mathbf{P_i} = \{p_1, p_2, \dots, p_m\}$ to be filled with predicate values. SQL templates alone cannot be executed directly by the execution engine.
\end{definition}

\example \label{ex:sql-template}
{Given the schema: \texttt{Users(user\_id, user\_name)} and \texttt{Orders(order\_id, user\_id, order\_amount)}, this template retrieves the ids of users who have orders that exceed a certain amount. There is \textcolor{blue}{a placeholder $p_1$} for this amount.}
\begin{lstlisting}
SELECT UNIQUE(user_id)                
FROM orders
WHERE orders.order_amount > ~{$\color{blue}{p_1}$}~;
\end{lstlisting}

\begin{definition}[SQL Query]
\label{definition:sql-query}
SQL queries are SQL statements that can be executed directly by the query execution engine. SQL templates $\mathbf{T}$ can be instantiated into executable SQL queries by replacing the placeholders $\mathbf{P}$ with predicate values. 
\end{definition}

\example \label{ex:sql-query}
{Given the SQL template used in Example~\ref{ex:sql-template}, we instantiate this template by replacing the placeholder for amount with \textcolor{blue}{a predicate value of 500}.}
\begin{lstlisting}
SELECT UNIQUE(user_id)                
FROM orders
WHERE orders.order_amount > ~500~;
\end{lstlisting}

\begin{definition}[Specifications for SQL Templates]
\label{definition:nl-specifications}
Specifications $\mathbf{S} = \{s_1, s_2, \dots, s_n\}$ control the structure and features of SQL templates. Each specification may include numerical constraints, such as limits on the number of tables, joins, or aggregations, as well as structural constraints, such as the presence of nested queries or the use of complex scalar expressions.

\end{definition}

\example \label{ex:nl-specification}
Users can provide a natural language specification, such as: ``I want a complex SQL template that accesses 30\% of the tables, includes 5 joins, and performs 3 aggregations.'' Alternatively, for business intelligence use cases, a user might specify: ``I want an SQL template with no joins but with complex scalar expressions,'' which is not supported by any existing benchmark~\cite{10.1145/3209950.3209952}.


\begin{definition}[Customized SQL Template Generation]
\label{definition: customized-templates}
Given a database $\mathbf{D}$ and a collection of user specifications $\mathbf{S}$, where $s_i$ is a natural language requirement on the SQL template such as ``This SQL template should have two joins and access three tables'', we want to generate a set of SQL templates $\mathbf{T}$ such that $t_i$, \textcolor{black}{after instantiation}, is executable on $\mathbf{D}$ and satisfies $s_i$. Each SQL template $t_i$ can be instantiated into SQL queries by replacing the placeholders $\mathbf{P}$ with predicate values. 
\end{definition}

\example \label{ex:customized-template}
{Given the schema used in Example~\ref{ex:sql-template}, the user wants an SQL template that includes at least \textcolor{blue}{one JOIN} and \textcolor{red}{a nested query with an aggregation.} This template has two placeholders $p_1$ and $p_2$ for two predicate values.}
\begin{lstlisting}
SELECT u.user_name, SUM(o.order_amount)                          
FROM users AS u
~JOIN orders AS o ON u.user_id = o.user_id~                    
WHERE u.user_id IN (                              
    -SELECT  user_id-
    -FROM    orders-
    -GROUP BY user_id-
    -HAVING  COUNT(order_id) > {$\color{red}{p_1}$} -
)
AND  o.order_amount >= {$p_2$};         
\end{lstlisting}

\begin{definition}[Correctness of SQL Templates.]
\label{definition: cost-query}
Correct SQL templates should satisfy two requirements. (1) SQL templates should not have syntax errors, and the instantiated SQL queries should be executable on the target database. (2) SQL templates should satisfy user-defined specifications. Note that since we are targeting generating SQL queries for testing and benchmarking purposes, we do not have requirements on the semantics of these templates.

\end{definition}

\begin{definition}[Cost-Aware Query Generation]
\label{definition: cost-query}
Given a database $\mathbf{D}$, an SQL template $t_i$ with placeholders $\mathbf{P_i}$, and a target query cost $\mathbf{C}$, we try to produce an SQL query by replacing the placeholders $\mathbf{P_i}$ with different predicate values, and identifying appropriate predicate values such that this query can have the target cost $\mathbf{C}$. The query cost type could be cardinality, execution plan cost, execution time, or any user-defined one. These cost metrics can be obtained by estimations from the query optimizer or by actual execution.
\end{definition}

\example
\normalfont
{Given the database schema used in Example \ref{ex:sql-template}, the user provides that \texttt{Users} has 1000 rows and \texttt{Orders} have 500 rows. The user wants one query that \textcolor{blue}{returns 200 rows (cardinality=200)} by \textcolor{red}{joining} the two tables with \textcolor{red}{an appropriate predicate condition.} By replacing the placeholder with the predicate value \textcolor{red}{50}, this query successfully returns 200 rows.}
\begin{lstlisting}
SELECT u.user_name, o.order_date
FROM users u 
-JOIN orders o ON u.user_id = o.user_id-
-WHERE o.order_id > 50; -
# Execute: succesfully returns ~200 rows~
\end{lstlisting}

\begin{definition}[Cost Distributions of SQL Queries.]
\label{definition: cost-query}
Each SQL query incurs a cost, and users can specify a target cost distribution that the generated queries are expected to follow. The Wasserstein Distance~\cite{wassersteindistance} can be used to measure the similarity between the user-specified target distribution and the cost distribution of the generated SQL queries. The system should generate SQL queries with costs that reduce this distance as much as possible within a given time budget. 
\end{definition}

\begin{figure}[h]
    \centering
    \includegraphics[width=0.72\linewidth]{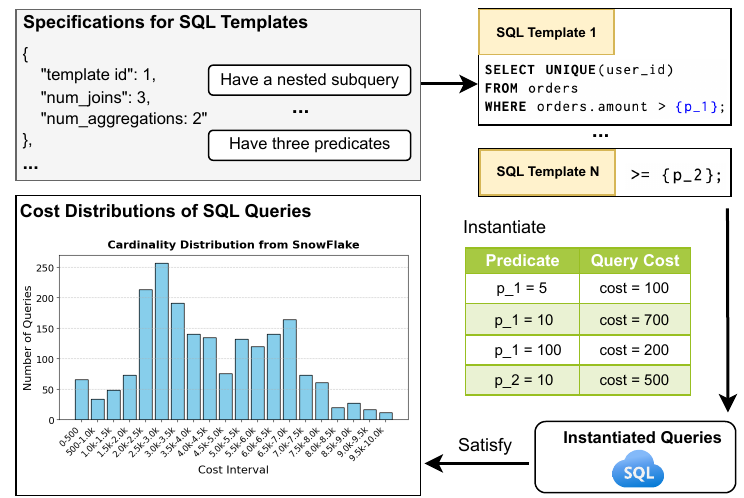}
    \caption{A Running Example of SQL Workload Generation}
    \label{fig:example}
\end{figure}

\begin{definition}[Customized and Realistic Workload Generation]
\label{definition: workload}
Given a database $\mathbf{D}$, a set of user specifications $\mathbf{S}$ for SQL templates, the number $\mathbf{N}$ of queries to generate, and a target cost distribution $d^*$, we want to generate an SQL workload where the underlying SQL templates satisfy $\mathbf{S}$ and the instantiated $\mathbf{N}$ queries match the target cost distribution $d^*$.
\end{definition}
\example
\normalfont Figure~\ref{fig:example} presents an end-to-end example of the customized and realistic SQL workload generation. On the left side, users provide specifications for SQL templates, including both JSON-formatted numerical constraints and natural language instructions. These specifications guide the generation of SQL templates. Each SQL template contains placeholders for predicate values. By substituting these placeholders with different predicate values, the system generates queries of varying costs. The costs of all generated queries, or a selected subset, are expected to match the user-specified target cost distribution. 

\begin{figure}[ht]
    \centering
    \includegraphics[width=0.77\linewidth]{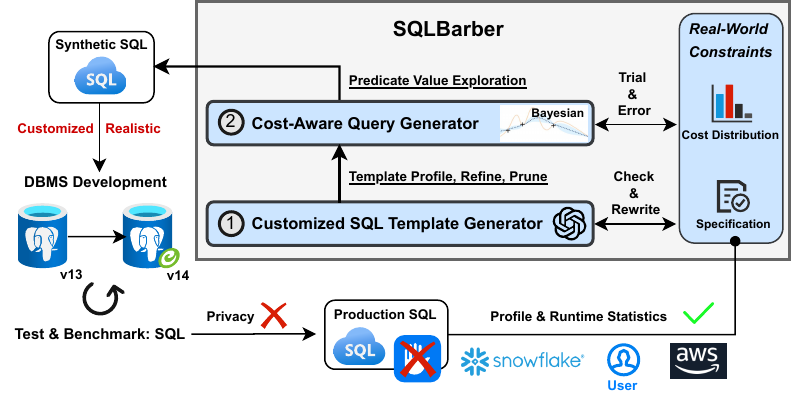}
    \caption{System Overview of SQLBarber}
    \label{fig: sqlbarber-overview}
\end{figure}

\section{System Overview}\label{sec:overview}


Figure~\ref{fig: sqlbarber-overview} shows the two components of SQLBarber: \ding{192} a customized SQL template generator and \ding{193} a cost-aware query generator. 

\noindent \ding{192} The customized SQL template generator uses LLMs to create SQL templates based on user-provided natural language specifications (Definition~\ref{definition:nl-specifications}). The specifications can define both the structure of the SQL template (e.g., the inclusion of nested subqueries) and its features (e.g., the number of tables accessed, the number of joins and aggregations, or the presence of certain operators such as groupby). First, the Template Generator connects to the target database and produces a text summary of the database schema and join paths. Subsequently, the Template Generator combines the user specification, the database schema, and the join paths to construct a prompt to invoke LLM to generate SQL templates. Due to the hallucination problem in LLMs (i.e., returning a response that is false or entirely made up), it is possible that the generated SQL templates have syntax errors or do not satisfy the user specifications. Since LLMs possess the capability to evaluate and self-correct their outputs, SQLBarber employs LLMs to reason about whether an SQL template satisfies user specifications and to identify the reasons behind it. If a template fails, LLM uses the reasoning results along with the error messages from the DBMS to iteratively rewrite this template, ensuring that it becomes both executable \textcolor{black}{after instantiation} and compliant with the user specification. 


\noindent \ding{193} The cost-aware query generator produces a large set of SQL queries that match any user-specified cost distribution. It follows a two-phase process that combines profiling with Bayesian Optimization (BO) to instantiate SQL templates using appropriate predicate values. First, the Query Generator profiles each template to estimate its capacity to produce queries across a range of costs. It replaces template placeholders with predicate values uniformly sampled from the target database, runs the resulting queries on the DBMS, and records metrics such as estimated cardinality and cost from the query optimizer. Using these observations, it refines existing SQL templates to create new templates that address uncovered cost ranges and discards templates that do not help cover the desired cost distribution. Third, the Query Generator applies BO to search the predicate value space, balancing the exploitation of predicate values already known to satisfy the cost targets with the exploration of unknown predicate values that may further improve coverage. Finally, the Query Generator outputs the SQL queries whose templates meet the user specifications and whose measured costs conform to the target distribution, making them suitable for realistic testing and benchmarking.

\section{Customized SQL Template Generator}\label{sec:template-generator}




\noindent \textbf{Step 1: Database Schema Summary.} The Template Generator connects to the target database to gather necessary contexts for template construction and cost-aware query generation. Specifically, it extracts three categories of metadata: table-level, column-level, and constraint-level. For table-level information, we record table names, table sizes, and the number of tuples in a table. Table names tell LLMs which relations a query may access, table sizes allow LLMs to estimate the execution costs of a query (e.g., scanning large tables would take longer than scanning small tables), and the tuple counts provide information on the upper bound of the SQL cardinality. For column-level information, we supply column names, data types, and the number of distinct values, which guide LLMs to select appropriate filter predicates based on selectivity. Finally, constraint-level information includes primary and foreign keys, which indicate valid join pairs, and index metadata, which shows which columns can support index scans or index joins.

\begin{algorithm}[b]
\SetAlgoLined
\LinesNumbered
\caption{Iterative Template Check and Rewrite}
\label{alg:template_check_rewrite}
\SetKwInput{Input}{Input}
\SetKwInput{Output}{Output}
\Input{SQL template $T$; User specifications $S$; Target database $\mathcal{D}$; LLM $\mathcal{M}$; Max iterations $k$.}
\Output{Valid SQL template $T^*$ satisfying specifications $S$.}

\For{$i \in 1..k$}{
    \tcp{\textcolor{blue}{Check specification compliance}}
    $\langle satisfied, violations \rangle \gets \mathcal{M}.\textsc{ValidateSemantics}(T)$\;
    \If{not $satisfied$}{
        \tcp{\textcolor{blue}{Fix semantic violations}}
        $T^* \gets \mathcal{M}.\textsc{FixSemantics}(T^*, S, violations)$\;
    }

    \tcp{\textcolor{blue}{Check executability}}
    $\langle executable, errors \rangle \gets \mathcal{D}.\textsc{ValidateSyntax}(T)$\;
    
    \If{not $executable$}{
        \tcp{\textcolor{blue}{Fix execution errors}}
        $T^* \gets \mathcal{M}.\textsc{FixExecution}(T, errors)$\;
    }

    \If{$satisfied$ and $exetuable$}{
        \Return{$T^*$}\;
    }
}

\Return{$T^*$}\;
\end{algorithm}

\noindent \textbf{Step 2: Join Path Generation.} The Template Generator generates all possible join paths based on the constraint-level information collected in Step 1. For each attempt to construct an SQL template, the Template Generator randomly selects a join path that satisfies the user-specified number of joins, 
and uses this join path to guide the SQL template construction. \textcolor{black}{Currently, SQLBarber selects join paths only for the main query. Extending this mechanism to handle nested queries would be an interesting direction for future work.}
\textcolor{black}{Selecting only one join path offers several advantages.}
\textcolor{black}{First, the random selection makes the template generation diverse enough to cover different join patterns and table combinations across many attempts.}   Second, this approach is cost-effective due to prompt compression: instead of including metadata for all tables and columns, only those involved in the sampled join path are included, which reduces the number of tokens processed by LLMs. Third, the method improves reliability by mitigating the challenges of long-context processing in LLMs~\cite{bai-etal-2024-longbench}, especially when the target database has many tables.

\noindent \textbf{Step 3: Customized Prompt Construction.} The Template Generator constructs the prompt for invoking LLMs by combining the database context from Step 1, the sampled join path from Step 2, and the user-provided specifications for SQL templates. These specifications can be expressed in various textual formats, including structured formats such as JSON and unstructured natural language descriptions. In addition to fully customized specifications, SQLBarber also supports using specifications derived from real-world workloads, such as those extracted from Amazon Redshift~\cite{redset-vldb}. Each specification defines key properties of an SQL template, including the number of accessed tables, joins, scans, and aggregations. Taken together, the specifications across all templates describe the high-level characteristics of a real-world workload.

\noindent \textbf{Step 4: SQL Template Generation.} The Template Generator uses the prompt constructed in Step 3 to invoke LLMs and generate SQL templates. These templates can be instantiated into executable queries by replacing placeholders with concrete predicate values. It is important to note that this generation process is not cost-aware, and it is unclear whether the resulting queries will align with the target cost distribution. Therefore, the generated SQL templates are treated as seed templates, which will be further refined into template variants that better match the desired cost distribution.

\noindent \textbf{Step 5: Template Check and Rewrite.} The Template Generator iteratively checks and corrects the SQL templates generated in Step 4 to ensure these templates have no syntax errors and fulfill the user-provided specifications, addressing potential hallucination issues associated with the LLM. Algorithm~\ref{alg:template_check_rewrite} summarizes this process. It accepts an initial SQL template $T$, user specifications $S$, the target database $\mathcal{D}$, an LLM $\mathcal{M}$, and a maximum number of iterations $k$ as input. The algorithm iterates until it either generates a valid template or reaches the maximum iteration limit (Line 2). Each iteration consists of two sequential validation phases:


\begin{figure}[h]
    \centering
    \includegraphics[width=0.77\linewidth]{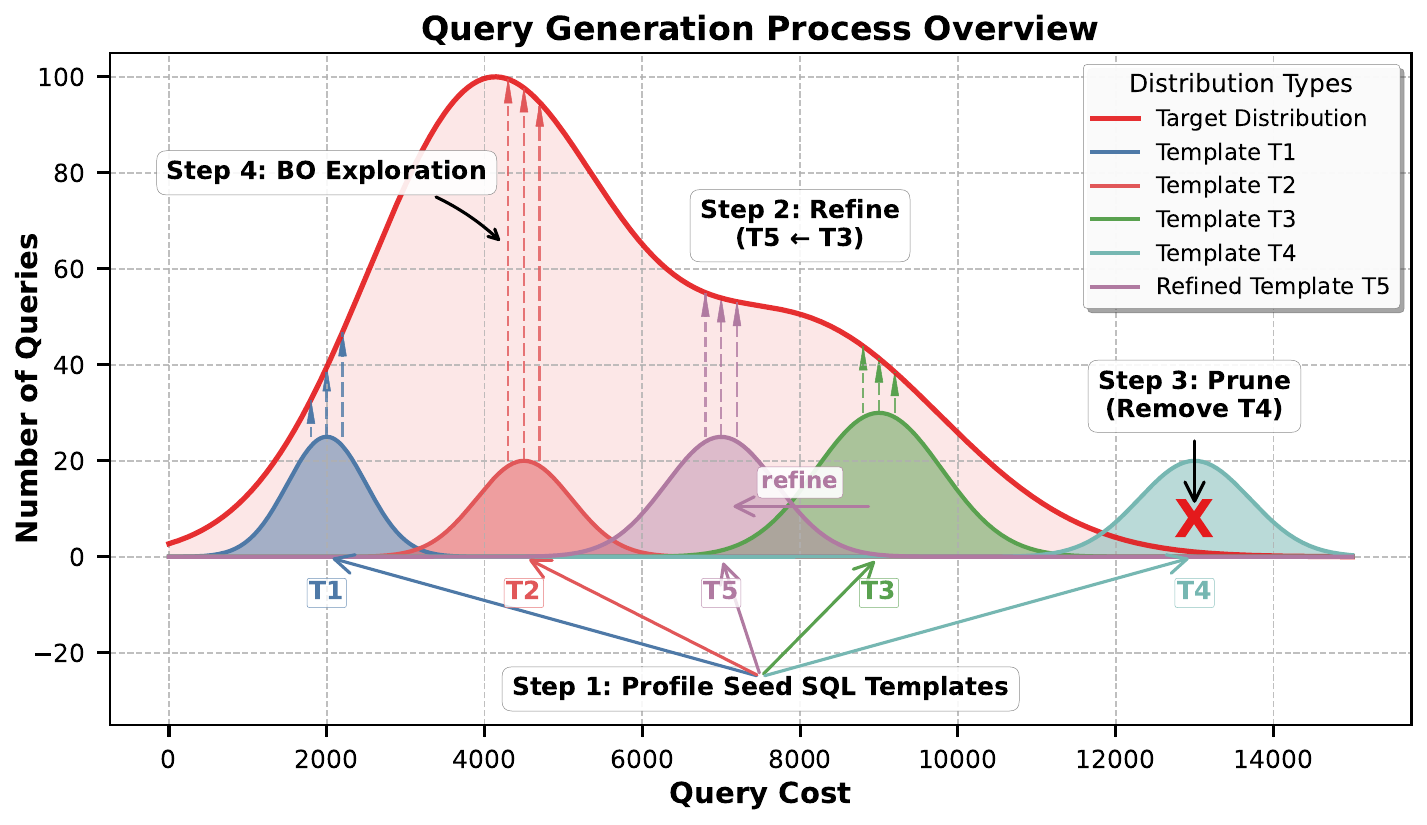}
    \caption{Cost-Aware Query Generation}
    \label{fig:query_generation}
\end{figure}

\textbf{Phase 1: Checking Specification Compliance (Lines 2–5).}
The algorithm first verifies if the current template $T^*$ meets the provided user specifications by invoking LLM’s \textsc{ValidateSemantics} function (Line 2). This function returns a boolean indicator $satisfied$ and a list of $violations$ indicating specification mismatches. If the template does not satisfy these specifications (Line 3), the LLM’s \textsc{FixSemantics} function generates a corrected template based on these violations (Line 4). The iteration then continues with the updated template (Line 5).

\textbf{Phase 2: Checking Database Executability (Lines 6–9).}
After passing the specification check, the algorithm validates the syntactic correctness of the template against the target database system by calling the \textsc{ValidateSyntax} function (Line 6). This function returns a boolean $executable$ flag and any syntax $errors$ encountered. If syntax errors are found (Line 7), the LLM's \textsc{FixExecution} function generates corrections based on the DBMS feedback (Line 8). Finally, we return the fixed SQL templates (Line 11).

\section{Cost-Aware Query Generator}\label{sec:query-generator}

Figure~\ref{fig:query_generation} illustrates the workflow of our cost-aware query generation process. The query generator takes as input the seed SQL templates produced by the customized SQL template generator (Section~\ref{sec:template-generator}). First, we profile these seed templates to evaluate their ability to generate queries across different cost levels (Section~\ref{subsec:template-profiling}). Based on the profiling results, we refine the templates to produce new ones that can better cover the target cost distribution and remove templates that do not contribute effectively (Section~\ref{subsec:template-refine-prune}). Finally, we apply Bayesian Optimization (BO) to search for predicate values that yield a sufficient number of queries in each cost range (Section~\ref{subsec:predicate-search}).

\subsection{Template Profiling via Strategic Sampling}\label{subsec:template-profiling}

SQLBarber profiles all SQL templates generated by the customized SQL template generator to assess their ability to produce queries across different cost levels. This profiling step is important for two reasons: (1) it identifies whether the existing SQL templates are sufficient to cover the entire cost range; if not, additional templates need to be generated, and (2) it determines which SQL templates are effective at generating queries in specific cost ranges, which in turn guides the optimization of the predicate search process. 

\textcolor{black}{
SQLBarber profiles SQL templates by instantiating them into executable SQL queries with different predicate values. It then evaluates these queries on the target database to collect performance metrics.} \textcolor{black}{Depending on the user-specified optimization target, these metrics may include estimated cardinality and execution cost obtained from the query optimizer using the \texttt{EXPLAIN} statement, or actual measurements such as execution time and resource usage (e.g., CPU and memory) collected by executing the queries. In our experiments, we follow prior work~\cite{learnedsqlgen, hillclimb, cost-aware-query-generation} to use \texttt{EXPLAIN} to get estimated cost and cardinality from the optimizer, and further extend it by using estimated CPU usage as the optimization target.}

\textcolor{black}{SQLBarber samples predicate values for each SQL template using a space-filling method, Latin Hypercube Sampling (LHS)~\cite{lhs}. Each SQL template may include multiple placeholders, with each placeholder corresponding to a column in the database schema. Instead of independently sampling values for each column, SQLBarber applies LHS to generate sample values that are more evenly distributed across the joint value space. This helps improve the coverage of the multi-dimensional space and reduces the risk of generating poorly distributed samples, which can occur with independent random sampling along each dimension.}

SQLBarber collects profiling results to support future optimizations. For each SQL template, it evaluates a small subset of sampled predicate values on the target database (e.g., 15\% of the total number of queries to be generated) to keep the profiling overhead low. As shown in Figure~\ref{fig:query_generation} (Step 1), this process produces a cost distribution for each SQL template, which is used to identify the cost ranges that the template can cover and to determine which templates are effective for generating queries within specific cost intervals.

\subsection{Adaptive Template Refinement and Pruning}\label{subsec:template-refine-prune}


SQLBarber adaptively refines existing SQL templates to generate new ones that cover cost ranges not represented by the original templates, and removes templates that do not contribute to the target distribution, enabling SQLBarber to meet arbitrary distributions.

As shown in Figure~\ref{fig:query_generation} (Step 2), given a target query cost range of 0 to 15,000, existing templates (T1 to T4) fail to produce queries with costs between 6,000 and 8,000. SQLBarber refines template T3 to create a new template T5 to generate queries within the 6,000 to 8,000 cost range. SQLBarber removes template T4, as it does not contribute effectively to the target cost distribution (Step 3).


SQLBarber employs a \textbf{cost-aware template refinement and pruning algorithm} to adapt initial SQL templates to better match the target cost distribution. First, it identifies low-coverage intervals, which are cost regions with fewer queries than the target requires. Second, it selects candidate templates for refinement based on their estimated utility in generating queries that fall into these low-coverage intervals. Third, it leverages LLMs to refine the most promising templates, producing new templates that are more likely to generate queries within the underrepresented intervals. Finally, it applies profiling to determine whether further refinement is necessary or if the current coverage is sufficient, in which case redundant templates can be pruned.


SQLBarber employs two processing strategies with different costs to process two types of underrepresented intervals: \emph{missing intervals}, which receive little or no coverage, and \emph{difficult intervals}, which fail to reach the target number of queries despite repeated refinements. This design is motivated by the observation that many missing intervals can be addressed efficiently by directly using LLMs to refine existing templates. In contrast, difficult intervals require a more costly strategy that leverages the in-context learning capabilities of LLMs. This involves iterative refinement based on historical trials and feedback from the DBMS. By applying lightweight processing to missing intervals and reserving expensive refinement for difficult intervals, SQLBarber achieves better cost-effectiveness. 

To distinguish underrepresented intervals, formally, let $\mathcal{I} = \{[l_1, u_1), [l_2, u_2), ..., [l_n, u_n)\}$ be the set of cost intervals \textcolor{black}{specified by the user (e.g, cardinality intervals $\{[0,1000), [1000,2000)\}$)}, and let $\mathcal{P} = \{(T_i, C_i)\}_{i=1}^{|\mathcal{T}|}$ denote the profiling results, where $T_i$ is an SQL template from $\mathcal{T}$ and $C_i = \{c_{i1}, c_{i2}, ..., c_{i|C_i|}\}$ is the set of observed execution costs of queries instantiated from $T_i$. We define the coverage of interval $j$ as the number of queries falling within it:
\begin{equation}\label{algo:coverage}
\scalebox{0.98}{$
c_j = \sum_{i=1}^{|\mathcal{T}|} |\{c \in C_i : l_j \leq c < u_j\}|
$}
\end{equation}



\begin{algorithm}[t]
\SetAlgoLined
\LinesNumbered
\caption{Cost-Aware Template Refinement \& Pruning}
\label{alg:template_refinement}
\SetKwInput{Input}{Input}
\SetKwInput{Output}{Output}
\Input{Templates $\mathcal{T} = \{T_i\}$; Profiling results $\mathcal{P} = \{(T_i, \mathbf{C}_i)\}$ where $\mathbf{C}_i$ is a cost vector; Cost intervals $\mathcal{I}$; Target distribution $\mathbf{d}^*$; LLM $\mathcal{M}$.}
\Output{Refined templates $\mathcal{T}^*$; Refined profiling results $\mathcal{P}^*$.}
\textbf{Initialize:} $\mathcal{T}^* \gets \mathcal{T}$, $\mathcal{P}^* \gets \mathcal{P}$, $\mathcal{H} \gets \{\}$ \tcp*{History $\mathcal{H}$: interval $\to$ list of (template, cost vector)}
\For{$(\tau, k, m, use\_hist) \in [(\tau_1, k_1, m_1, \text{false}), (\tau_2, k_2, m_2, \text{true})]$}{
    \For{$iter = 1$ \textbf{to} $k$}{
        $\mathbf{c} \gets$ \textbf{array}($|\mathcal{I}|$, 0) \tcp*{\textcolor{blue}{Coverage vector}}

        $\mathbf{c}[j] = \sum_{i=1}^{|\mathcal{T}|} |\{c \in C_i : l(\mathcal{I}_j) \leq c < u(\mathcal{I}_j)\}|$
        
        $\mathcal{I}_{low} \gets \{j : \mathbf{c}[j] < \tau \cdot d^*_j\}$ \tcp*{\textcolor{blue}{Low coverage}}
        
            $\mathcal{N} \gets$ \textcolor{black}{\textit{RefineForIntervals}}~($\mathcal{I}_{low}$, $m$, $use\_hist$)\;
        $\mathcal{T}^* \gets \mathcal{T}^* \cup \mathcal{N}$\;
    }
}
\Return{$\langle \mathcal{T}^*, \mathcal{P}^*\rangle$}\;
\SetKwProg{Fn}{Function}{:}{}
\vspace{0.3cm}
\Fn{\textcolor{black}{\textit{RefineForIntervals}}~($\mathcal{I}_{target}$, $m$, $use\_history$)}{
        $\mathcal{N} \gets \emptyset$  \tcp*{\textcolor{blue}{New templates}}
    
    \For{each interval $j \in \mathcal{I}_{target}$}{
        $\mathbf{s}_j \gets$ ComputeCloseness($\mathcal{P}^*$, $j$) \tcp*{\textcolor{blue}{Eq.(\ref{algo:closeness})}}
        $\mathcal{T}_j \gets$ SelectTopK($\mathbf{s}_j$, $m$)\;
        
        \For{each template $T \in \mathcal{T}_j$}{
            $history \gets \text{Null}$\;
            \If{$use\_history$ \textbf{and} $j \in \mathcal{H}$ \textbf{and} $|\mathcal{H}[j]| > 0$}{
                $history \gets \mathcal{H}[j]$\;
            }
            
            $T_{new} \gets \mathcal{M}.$RefineTemplate($T$, $\mathbf{C}_T$, $\mathcal{I}_j$, $history$)\;
            
            $\mathbf{C}_{new} \gets$ Profile($T_{new}$) \tcp*{\textcolor{blue}{Cost vector}}
            
            \If{Not Prune($\mathbf{C}_{new}$, $\mathcal{I}_{target}$, $\mathbf{d}^*$) \tcp{\textcolor{blue}{Eq.(\ref{algo:prune})}}}{
                $\mathcal{N} \gets \mathcal{N} \cup \{T_{new}\}$\;
                $\mathcal{P}^* \gets \mathcal{P}^* \cup \{(T_{new}, \mathbf{C}_{new})\}$\;
                $\mathcal{H}[j] \gets \mathcal{H}[j] \cup \{(T_{new}, \mathbf{C}_{new})\}$ 
            }
        }
    }
    
    \Return{$\mathcal{N}$}\;
}
\end{algorithm}

To estimate the utility of an SQL template $T_i$ to produce queries in a given cost interval $j$, we define a \emph{Closeness} metric as follows:
\begin{equation}\label{algo:closeness}
s_{ij} = \frac{1}{1 + \bar{d}_{ij}} \cdot v_i
\end{equation}
where $\bar{d}_{ij}$ is the average distance between the template's historical query costs and interval $j$, \textcolor{black}{and $v_i = \frac{|\text{unique}(C_i)|}{|C_i|}$, ranging from $\frac{1}{|C_i|}$ to 1, represents the ratio of distinct cost values to total queries, penalizing templates with limited cost diversity. The average distance is calculated as follows:}
\begin{equation}
\scalebox{0.98}{$
\bar{d}_{ij} = \frac{1}{|C_i|} \sum_{q \in C_i} \text{dist}(\text{cost}(q), [l_j, u_j])
$}
\end{equation}
$\text{dist}(c, [l, u])$ returns 0 if $c \in [l, u]$, $(l - c)$ if $c < l$, and $(c - u)$ if $c > u$.  This closeness score balances proximity to the target interval with the ability to generate diverse costs.

We use profiling to determine whether a newly refined template should be accepted. A refined template $T_{new}$ with profiling results $C_{new}$ is accepted if it satisfies the following condition:
\begin{equation}\label{algo:prune}
\scalebox{0.9}{$
\text{Prune}(T_{new}) = \begin{cases}
\text{false} & \text{if } \exists j \in \mathcal{I}_{target}: |\{c \in C_{new} : c \in \mathcal{I}_j\}| > 0 \\
\text{false} & \text{if } D(\mathbf{d}^c + \mathbf{v}_{new}, \mathbf{d}^*) < D(\mathbf{d}^c, \mathbf{d}^*) \\
\text{true} & \text{otherwise}
\end{cases}
$}
\end{equation}
where $\mathcal{I}_{target}$ represents the set of underrepresented intervals, $\mathbf{d}^c$ is the current distribution, $\mathbf{v}_{new}$ is the distribution contribution from $C_{new}$, and $D(\cdot, \cdot)$ is a distance metric (e.g., Wasserstein
Distance) between distributions. \textcolor{black}{Templates that do not generate queries within underrepresented intervals (do not satisfy the first condition in Eq.~\ref{algo:prune}) or fail to improve the overall distribution coverage (do not satisfy the second condition in Eq.~\ref{algo:prune}) are pruned.}

The workflow is outlined in Algorithm~\ref{alg:template_refinement}. It takes as input the initial templates $\mathcal{T}$, profiling results $\mathcal{P}$, cost intervals $\mathcal{I}$, target distribution $\mathbf{d}^*$, and an LLM $\mathcal{M}$. The algorithm outputs both the refined templates $\mathcal{T}^*$ and their corresponding profiling results $\mathcal{P}^*$. The algorithm employs a two-phase iterative refinement strategy (Lines 2-11). In each phase, it computes the coverage vector $\mathbf{c}$ (Lines 4-5), where $\mathbf{c}[j]$ counts the number of cost values from all templates that fall within interval $\mathcal{I}_j$. Intervals with low coverage are identified as missing or difficult intervals based on the threshold $\tau \cdot d^*_j$ and based on the current phase (Line 2 and Line 6).
The core refinement logic is encapsulated in the \textsc{RefineForIntervals} function (Lines 12-32). For each low-coverage interval $j$, the function computes closeness scores $\mathbf{s}_j$ using Equation~(\ref{algo:closeness}) (Line 15) and selects the top-$m$ templates with the highest scores (Line 16). Each selected template $T$ is then refined by the LLM (Line 22), which generates a new template $T_{new}$ targeted at the specific interval $\mathcal{I}_j$. 

The two phases differ in their parameters and use of historical context. The first phase uses parameters $(\tau_1=0.2, k_1=3, m_1=3)$ without history, performing standard refinement for missing intervals. The second phase employs stricter parameters $(\tau_2=0.1, k_2=5, m_2=5)$ with history enabled, targeting persistently difficult intervals. When history is enabled (Lines 19-22), the LLM receives both the current template and previous refinement attempts from $\mathcal{H}[j]$, enabling few-shot learning from past trials. This design is to optimize the trade-off between quality and cost. The parameter values are selected through simple tuning and have shown stable performance across ten benchmarks.

\textcolor{black}{SQLBarber guides LLMs to refine existing SQL templates by leveraging both the historical cost ranges of these templates and the target cost range. It provides hints for three types of refinement operations in the prompt: (1) selecting alternative join paths that access larger or smaller tables to adjust the query cost, (2) modifying the SQL structure by adding or removing predicate conditions or changing filtering columns according to their selectivity, and (3) allowing the LLMs to refine templates freely without explicit constraints. The LLM then decides which operations to apply based on the current SQL template, its cost range, the target cost range, and the available table and column information, and outputs one or more refined SQL templates as candidates.}

Each newly generated template undergoes profiling \textcolor{black}{(using the same profiling method in Section~\ref{subsec:template-profiling})} to obtain its cost vector $\mathbf{C}_{new}$ (Line 23) and is evaluated against the pruning criteria (Line 24). Templates that pass the pruning check—either by filling underrepresented intervals or reducing the overall distribution distance as defined in Equation~(\ref{algo:prune})—are added to the refined template set $\mathcal{N}$ and their profiling results are recorded in $\mathcal{P}^*$ (Lines 26-27). The history $\mathcal{H}$ is also updated to track all refinement attempts (Line 28). This pruning ensures comprehensive generation across the entire cost range while minimizing unnecessary costs.

\subsection{Predicate Search via Bayesian Optimization}\label{subsec:predicate-search}


In Figure~\ref{fig:query_generation}, after template generation, refinement, and pruning (Steps 1–3), the cost range is fully covered along the horizontal dimension. However, the number of queries per cost interval (the vertical dimension) remains insufficient. SQLBarber uses Bayesian Optimization (BO) to search the predicate values of SQL templates, aiming to produce enough queries within each interval (Step 4).

\begin{algorithm}[htbp]
\SetAlgoLined
\LinesNumbered
\caption{Adaptive BO-Based Predicate Search}
\label{algo:bo-predicate-search}
\SetKwInput{Input}{Input}
\SetKwInput{Output}{Output}
\Input{Templates $\mathcal{T}$, Profiling data $\mathcal{P}$, Number of templates to explore in one iteration $K$, Target distribution $\mathbf{d}^*$, Current distribution $\mathbf{d}$}
\Output{Updated distribution $\mathbf{d}'$, Generated queries $\mathcal{Q}$}

\tcp{\textcolor{blue}{Main loop: iteratively close distribution gaps}}
\While{$\exists j : \mathbf{d}^*[j] > \mathbf{d}[j]$}{
    \tcp{\textcolor{blue}{Step 1: Identify the most critical gap}}
    $j^* \gets \arg\max_j \{\mathbf{d}^*[j] - \mathbf{d}[j]\}$\;
    
    \tcp{\textcolor{blue}{Step 2: Select most promising templates}}
    $\mathcal{T}_{\text{viable}} \gets \{T \in \mathcal{T} :
        \text{HasSufficientSearchSpace}(T) \land
        \text{HasDiverseCost}(T) \land
        \text{NotPreviouslyIneffective}(T, j)\}$\;

    $\mathcal{T}_{cand} \gets\text{TopK}_{T \in \mathcal{T}} \left( \text{ClosenessScore}(T, j, \mathcal{P}),\, K \right)$\;
    
    
    \If{$\mathcal{T}_{cand} = \emptyset$}{
        Mark $j^*$ as exhausted and continue\;
    }
    
    \tcp{\textcolor{blue}{Step 3: Explore predicates via BO}}
    \For{each template $T \in \mathcal{T}_{cand}$}{
        $\mathcal{Q}_{new} \gets$ \textsc{BayesianOptimize}($T, j^*, \mathbf{d}^*[j^*] - \mathbf{d}[j^*]$)\;
        
        \tcp{\textcolor{blue}{Step 4: Update and evaluate progress}}
        $\mathbf{d}' \gets$ \textsc{UpdateDistribution}($\mathbf{d}, \mathcal{Q}_{new}$)\;
        
        \If{insufficient improvement}{
            Mark $(j^*, T)$ as ineffective\;
        }
        
        $\mathbf{d} \gets \mathbf{d}'$\;
        $\mathcal{Q} \gets \mathcal{Q} \cup \mathcal{Q}_{new}$\;
    }
}
\Return{$\langle\mathbf{d}, \mathcal{Q}\rangle$}\;


    
\end{algorithm}

SQLBarber employs an \textbf{Adaptive BO-Based Predicate Search} algorithm to systematically fill gaps between the current and target query cost distributions. As shown in Algorithm~\ref{algo:bo-predicate-search}, it follows a four-step iterative process. First, it identifies the interval $j^*$ with the largest gap between target and current distributions (Line 2). Second, it selects the most promising templates using multiple selection criteria (Line 3-4). Third, for each selected template, it explores the predicate value space using BO (Line 7-8). Finally, it updates the distribution and evaluates progress to adaptively adjust future exploration based on historical performance (Lines 9–14).

The algorithm selects promising SQL templates based on three key criteria (Lines 3-4) and uses closeness-based sampling to balance effectiveness with efficiency (Line 5). It filters for viable templates by checking: (1) sufficient remaining search space to generate enough queries for the gap size, (2) different predicate values can produce queries with diverse costs to avoid redundant explorations, and (3) exclusion of previously ineffective $\langle template, interval \rangle$ pairs (Line 3). Then it performs weighted sampling based on closeness scores (Eq.~(\ref{algo:closeness})) to select Top-$K$ templates (e.g., 10) (Line 4), balancing effectiveness with computational efficiency.

The optimization phase leverages BO to efficiently explore predicate values for each selected SQL template. The BO budget is allocated proportionally to the gap size ($\mathbf{d}^*[j^*] - \mathbf{d}[j^*]$, Line 9), ensuring computational resources are allocated based on need. BO iteratively explores the predicate space by balancing exploitation of values meeting cost constraints with exploration of uncertain regions. The objective is to minimize the distance between the actual cost $c$ of an instantiated query and the target cost interval $[c_l, c_r]$:
\begin{equation}
\scalebox{0.95}{$f(c) = \begin{cases}
0, & \text{if } c_l \leq c \leq c_r \\
1 - \max\left(\min\left(\frac{c}{c_l}, \frac{c_l}{c}\right), \min\left(\frac{c}{c_r}, \frac{c_r}{c}\right)\right), & \text{otherwise}
\end{cases}$}
\end{equation}
We use Random Forest as the surrogate model and Expected Improvement as the acquisition function~\cite{lindauer-jmlr22a} to guide the search in the high-dimensional predicate space, with historical optimization runs reused to warm-start the model when applicable.

The algorithm employs adaptive learning to improve efficiency. After generating queries, it evaluates whether the produced queries yield sufficient improvement (Line 11). If a template fails to generate queries that improve the target interval, it is marked as ineffective for that interval (Line 12), preventing unnecessary future attempts. When no templates can improve a given interval, the interval is marked as exhausted (Lines 5–7), allowing the algorithm to concentrate on intervals with achievable optimization potential. This process continues until the generated cost distribution closely matches the target or until no further improvable intervals remain. The algorithm’s adaptive behavior supports efficient resource usage and handles cases where some cost ranges cannot be improved using the current set of templates.

\begin{figure*}[th!]
    \centering
    \includegraphics[width=1\linewidth]{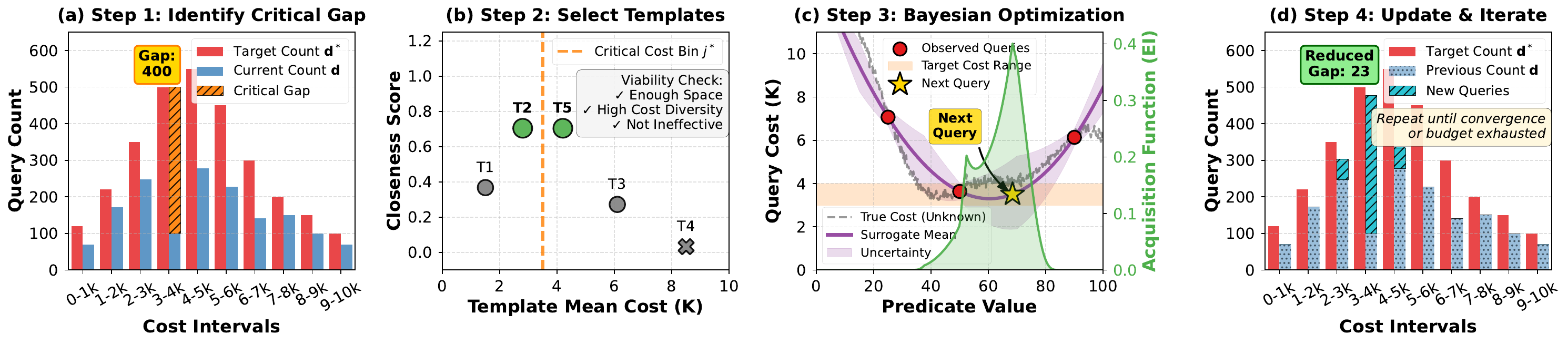}

    \caption{A Running Example of the Adaptive BO-Based Predicate Search Algorithm}
    
    \label{fig:example-predicate-search}
\end{figure*}

\example
\normalfont Figure~\ref{fig:example-predicate-search} presents a running example of Algorithm~\ref{algo:bo-predicate-search}. First, the cost range $[3k\text{--}4k]$ lacks 400 queries, which we identify as the most critical gap. Second, the algorithm selects the most promising templates for generating queries in this interval. It applies three selection criteria to filter viable templates (T4 is excluded) and samples from the remaining ones based on their closeness to the gap (T2 and T5 are selected). Third, BO is used to explore the predicate value space of T2 and T5. It leverages observed queries, the surrogate model's predictions, and uncertainty estimates to choose the next predicate value by maximizing the Expected Improvement acquisition function. Finally, after exploring and evaluating predicate values, the newly generated queries fill the gaps, reducing the largest gap from 400 to 23 queries. This loop continues until convergence is reached or the computational budget is exhausted.

\section{Experimental Evaluation}\label{sec:experi}








\subsection{Experimental Setup}

\noindent \textbf{Datasets.} Following prior works~\cite{learnedsqlgen, blueprinting}, we use two datasets as the underlying databases. (1) TPC-H~\cite{tpc-h-2.17.1} is a widely used benchmark to model a business analytics scenario with eight tables. We set the scale factor to $10$. 
(2) IMDB~\cite{job} is a real-world dataset of 21 tables that contains information related to films and television programs. 


\begin{figure*}[htbp]
    \centering
    \includegraphics[width=1\linewidth]{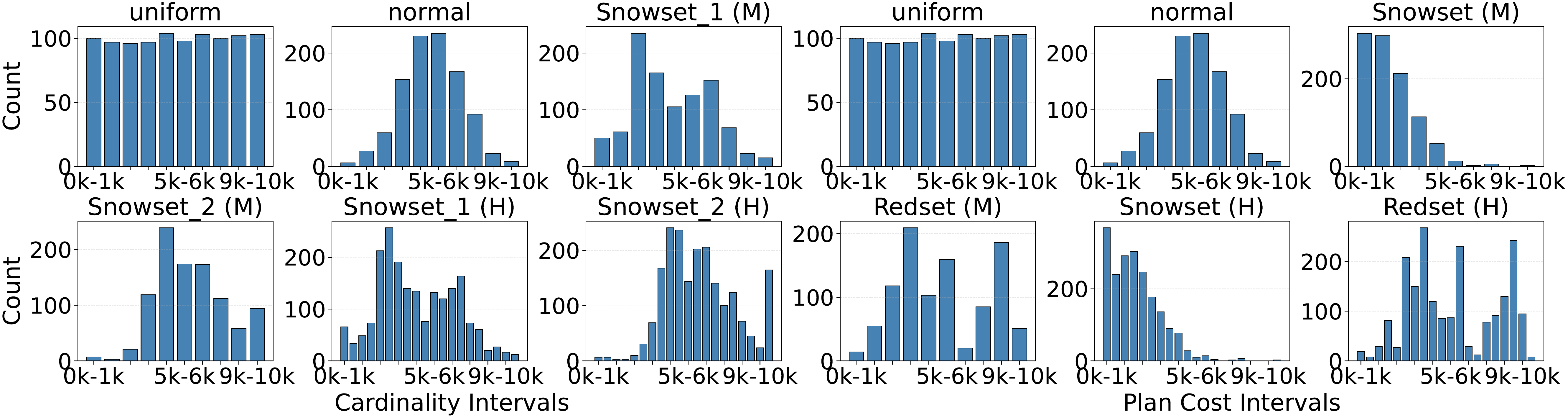}
    
    \caption{\textcolor{black}{Target Query Distributions (Left: Cardinality, Right: Execution Plan Cost)}}
    \label{fig:target_distributions}
\end{figure*}

\noindent \textbf{Benchmarks.} \textcolor{black}{As shown in Figure~\ref{fig:target_distributions}}, we create and open-source ten benchmarks for SQL workload generation.
The uniform distribution measures the average ability of a method to generate queries in any given cost range, and the normal distribution simulates the distributions of existing benchmarks such as TPC-H and TPC-DS \cite{redset-vldb}. The eight real-world distributions are derived from the execution statistics provided by Snowflake \cite{snowflake-nsdi20} and Redshift \cite{redset-vldb}. 
For the uniform, normal, and real-world distributions with medium difficulty (denoted as ``M''), we generate 1,000 queries across 10 cost intervals. For the more difficult distributions (denoted as ``H''), we generate 2,000 queries across 20 cost intervals. \textcolor{black}{Similarly to prior work~\cite{learnedsqlgen}, we set the target cost range to $[0, 10k]$ and actual values are linearly mapped to this range.} For SQL template specifications, we use a randomly selected workload from Amazon Redshift~\cite{redset-vldb}, which contains 28 tables and 24 SQL templates. \textcolor{black}{Each} template is annotated with the attributes \textit{num\_tables\_accessed}, \textit{num\_joins}, and \textit{num\_aggregations}. We also construct three natural language instructions to control (1) the presence of a nested subquery, (2) the number of predicate values, and (3) the use of the GROUP BY operator. Each template is randomly assigned at least one of these instructions.

\noindent \textbf{Target Cost Metrics.} 
\textcolor{black}{Following prior studies~\cite{learnedsqlgen, cost-aware-query-generation, hillclimb}, we focus on two types of SQL costs estimated by the query optimizer: (1) \texttt{Cost}, which represents the optimizer’s estimated expense of executing a query, and (2) \texttt{Cardinality}, which is the number of rows returned by the query. SQLBarber also supports other cost functions beyond these two metrics, such as CPU usage, memory usage, and runtime. We demonstrate this capability by designing a CPU-based cost function that targets a real-world CPU usage distribution from Snowflake~\cite{snowflake-nsdi20} in Section~\ref{subsec: further-anlaysis}.} 




\noindent \textbf{Evaluation Metrics.} We use three metrics to evaluate SQL generation methods. (1) {End-to-End Query Generation Time:} This metric measures the total time required to generate a specified number of SQL queries (e.g., $1k$) that satisfy user-defined constraints. These constraints include the cost type (e.g., cardinality or execution cost), the target cost distribution (synthetic or real-world), the specified cost range (e.g., $[0, 10k]$), and the number of cost intervals ($10$ or $20$ intervals). (2) {Wasserstein Distance (Earth Mover's Distance~\cite{panaretos2019statistical}):} This metric quantifies the difference between the target distribution provided by the user and the distribution of the generated queries. A smaller value indicates a closer match to the target, making it a useful measure for evaluating the quality of the generated queries. (3) {Template Generation Accuracy:} This metric measures the percentage of generated SQL templates that are executable and align with the user-specified template constraints. We report this metric only for SQLBarber in the ablation study, as it is the only method that supports template generation from natural language.




\noindent \textbf{Baselines.} We compare SQLBarber with other state-of-the-art SQL generation methods. (1) LearnedSQLGen~\cite{learnedsqlgen} is a method that uses Reinforcement Learning (RL) to explore different SQL templates and predicate values to produce queries satisfying given cost constraints. (2) HillClimbing~\cite{hillclimb} is an approach that takes SQL templates as inputs, and uses heuristics to greedily tweak the predicate values in the given SQL templates to generate satisfied queries. We prepare about $16000$ SQL templates as inputs by randomly adding or removing predicates in the SQL templates provided by the benchmarks, the same approach used in \cite{learnedsqlgen}. Since these two baseline methods can generate queries for only one cost range per iteration, and the order of generating queries for different ranges may affect performance, we apply two heuristics to each baseline: \textbf{Order}, which generates queries from the lowest to the highest cost ranges; and \textbf{Priority}, which generates queries for the cost range with the largest number of missing queries in each iteration. The number of optimization iterations is equal to the number of cost intervals in the benchmark. We set a time budget of one hour for each optimization iteration. \textcolor{black}{(3) SQLStorm~\cite{sqlstorm} directly prompts LLMs to generate diverse SQL queries by setting the model temperature to 1. We use the official implementation and follow the original experimental setup, employing \texttt{gpt-4o-mini} to generate 5,000 SQL queries for each of seven prompts through the OpenAI batch processing service. After each batch, we estimate query costs using the same procedure as other baselines and compute the Wasserstein Distance. Generation stops once the distance reaches 0 or all seven prompts have been used. For a fair comparison, we omit the steps for cross-DBMS template compatibility to reduce runtime and cost, as this work focuses on PostgreSQL.}


\noindent \textbf{Implementation.} All methods use PostgreSQL v14.9. We implement SQLBarber in Python, and use OpenAI completion API to invoke the ``o3-mini'' model for the generation of SQL templates. We implement BO with the SMAC3 library \cite{lindauer-jmlr22a}. 

\noindent \textbf{Hardwares.} All experiments are conducted on a Ubuntu server with two Intel Xeon Gold 5218 CPUs (2.3GHz with 32 physical cores), 384 GB of RAM, and two GeForce RTX 2080Ti GPUs. Only LearnedSQLGen~\cite{learnedsqlgen} uses GPUs to train the neural networks used in RL. This is the same GPU type as reported by the authors.

\begin{figure*}[htbp]
    \centering
    \includegraphics[width=1\linewidth]{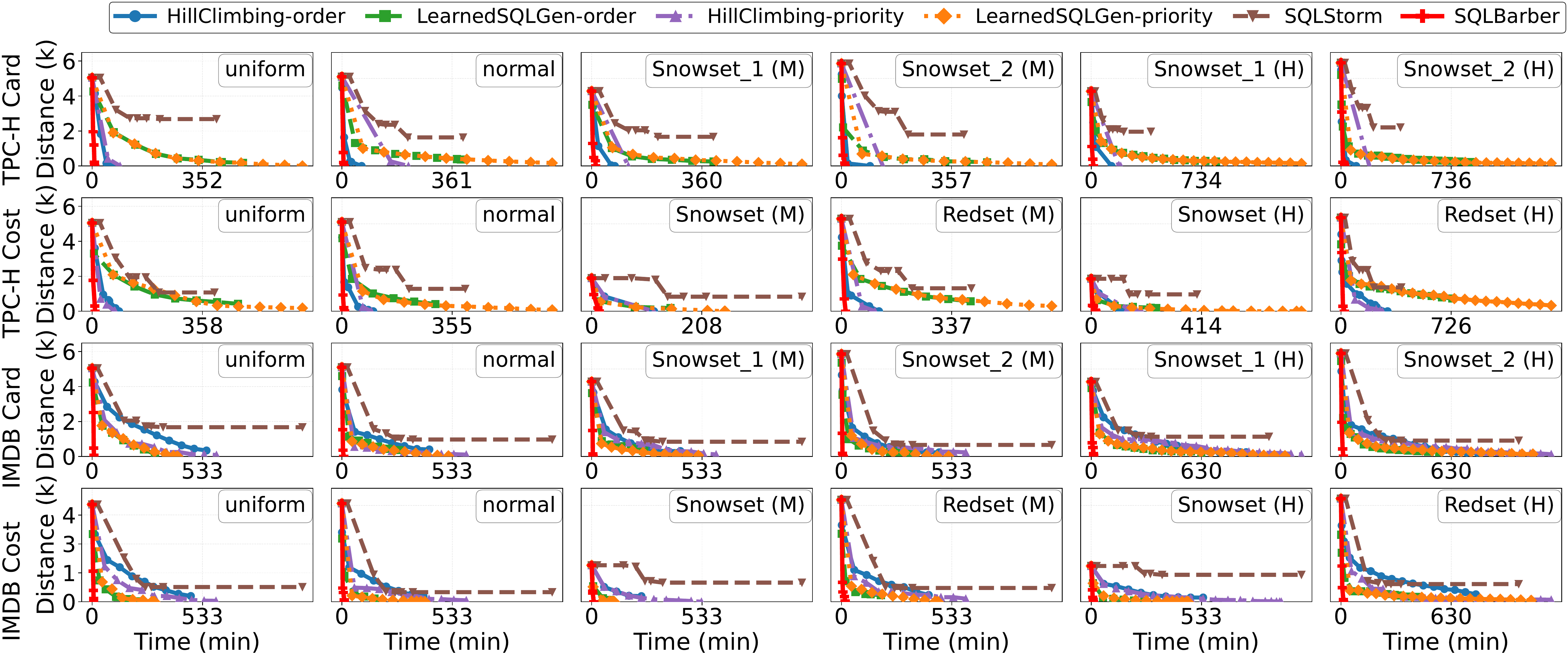}
    \caption{\textcolor{black}{Distance Between the Target Distribution and the Generated Distribution Over Time}}
    \label{fig:distance-over-time}
\end{figure*}

\begin{figure*}[htbp]
    \centering
    \includegraphics[width=1\linewidth]{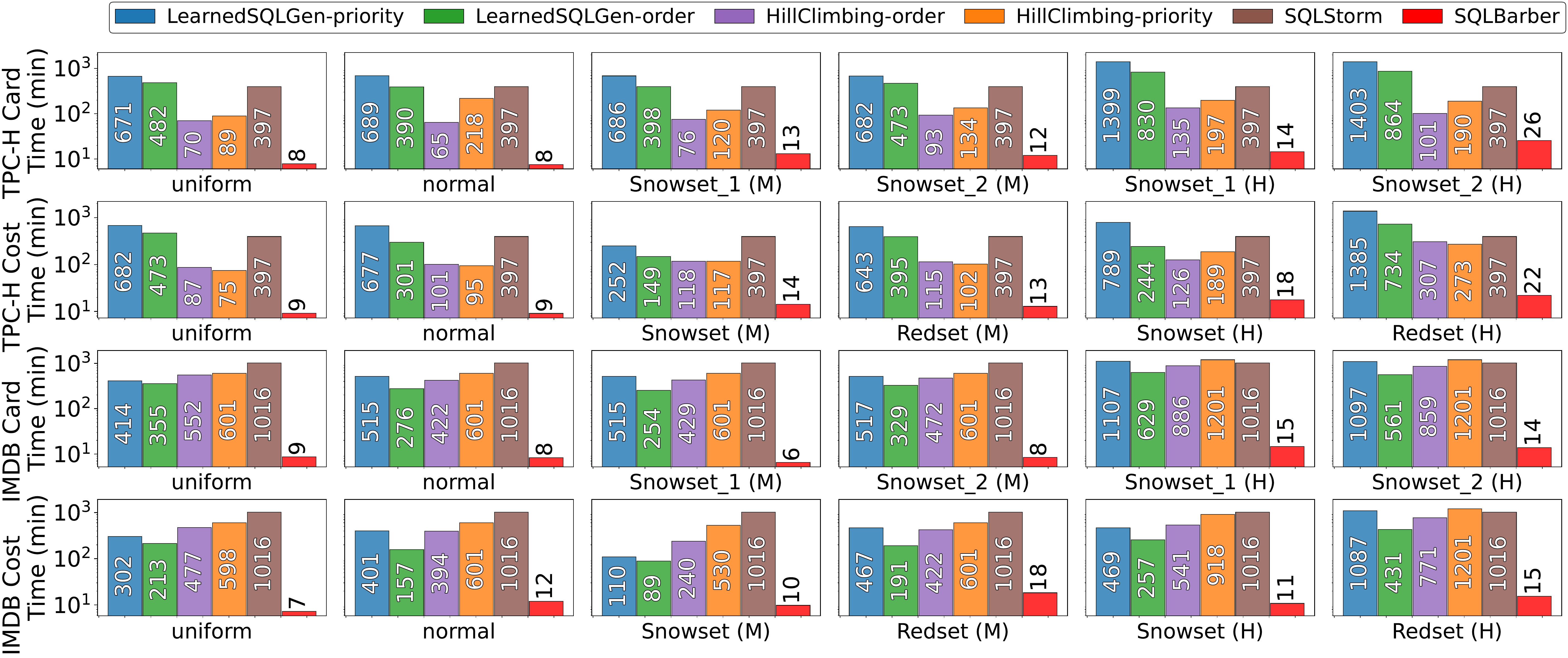}
    \caption{\textcolor{black}{SQL Workload Generation Time}}
    \label{fig:generation-time}
\end{figure*}

\subsection{Performance Study} \label{sec:performance}

Figure~\ref{fig:distance-over-time} and Figure~\ref{fig:generation-time} compares the effectiveness and efficiency of different SQL generation methods across all benchmarks, focusing on query cardinality (left) and execution plan cost (right). Each figure consists of three sub-figures arranged in rows for each benchmark: one distribution plot showing the target cost distribution, and two performance plots for the TPC-H and IMDB databases. Each performance plot includes (left) the Wasserstein distance between the target cost distribution and the cost distribution of the generated queries over time, and (right) the end-to-end query generation time.

SQLBarber significantly outperforms baseline methods in both the quality of the generated queries and the generation time. As shown in Figure~\ref{fig:distance-over-time} and Figure~\ref{fig:generation-time}, SQLBarber consistently reduces the distance to zero across ten benchmarks, two underlying databases, and two optimization targets, with an average generation time of about 12 minutes. In contrast, baseline methods typically fail to reduce the distance to zero, often resulting in distances between 500 and 2000, even after hundreds or thousands of minutes. This gap is more pronounced on harder benchmarks. For instance, on the “Snowset\_Card\_2\_Hard” benchmark using IMDB as the underlying database, baseline methods require between 561 and 1201 minutes to generate queries with a similar cost distribution. In comparison, SQLBarber generates queries with an exact cost distribution match (i.e., distance of zero) in only 14 minutes. In summary, SQLBarber achieves exact cost distribution matches with one to two orders of magnitude less generation time, while baseline methods fail to reach a zero distance despite extensive computation.

SQLBarber achieves substantial performance gains by adapting to different cost distributions through the use of refined SQL templates and an efficient predicate value search algorithm. In contrast, LearnedSQLGen~\cite{learnedsqlgen} requires a large number of samples for RL to capture the relationship among query cost, SQL templates, and predicate values, while HillClimbing~\cite{hillclimb} is limited by the quality of the input SQL templates. \textcolor{black}{An interesting observation is that the baseline methods exhibit different trade-offs between generation time and query quality, depending on the heuristic used. The “order” heuristic results in shorter generation time but larger distances. This difference arises because the “priority” strategy selects cost intervals with the most missing queries, allowing important yet under-covered intervals to be revisited. Revisiting these intervals improves performance, as additional generation attempts are allocated to intervals with many missing queries, rather than discarding them after a single attempt as in the “order” heuristic.}


\textcolor{black}{Finally, we compare SQLStorm, the only non–cost-aware baseline, with other methods. SQLStorm takes 394 and 1016 minutes to generate 35,000 queries for the TPC-H and IMDB databases, respectively. These times are shared across all distributions because queries are generated once, and each batch generation for one prompt is treated as a step in the distance-over-time plot (up to seven steps, corresponding to seven prompts). The long generation time results from its batch processing, which exchanges runtime for lower query costs (a 50\% discount from OpenAI). SQLStorm cannot reduce the distance to zero for any distribution, with final distances between 1000 and 3000.  Failing to reduce the distance to zero is expected since SQLStorm aims for query diversity rather than cost alignment. Note that, as shown in Figure~\ref{fig:query-comparison} (Section~\ref{subsec: further-anlaysis}), SQLStorm indeed produces the most diverse queries among all compared methods and existing benchmarks (e.g., covering the largest query length range). These results demonstrate that diversity-oriented query generation and realistic, constraint-aware query generation address distinct yet complementary problem settings, both of which are essential for DBMS development.}

\subsection{Scalability Study} \label{sec:scalability}

\begin{figure}[h]
    \centering
    \includegraphics[width=0.62\linewidth]{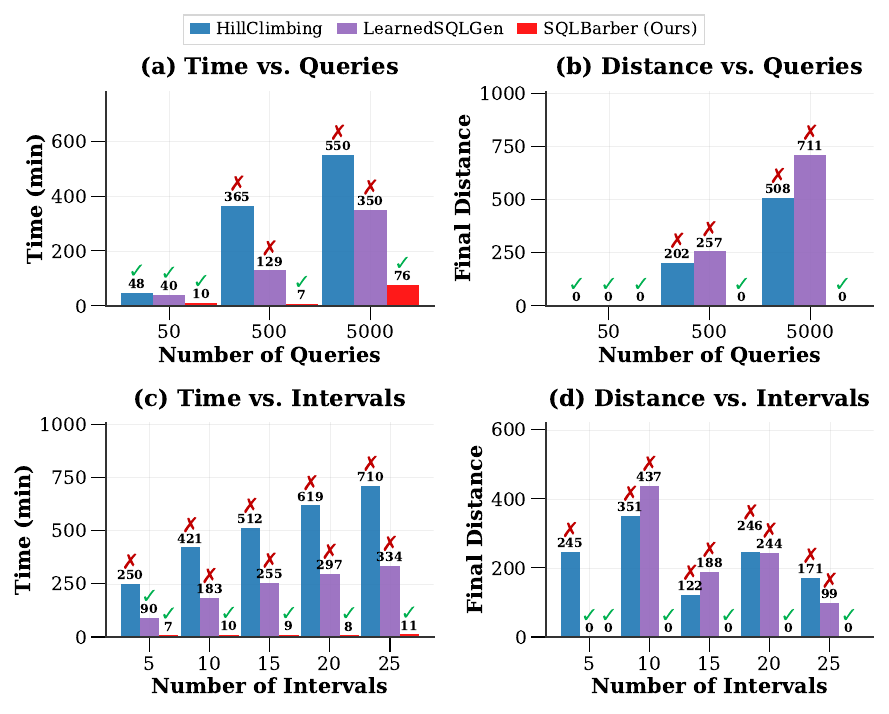}
    \caption{\textcolor{black}{Scalability Study on IMDB (Execution Plan Cost)}}
    \label{fig:scalability}
\end{figure}

Figure~\ref{fig:scalability} compares the scalability of different methods. A green check mark indicates that the method successfully reduces the distance to zero, whereas a red check mark indicates the failure.

\noindent \textbf{Scaling with the number of queries.} We evaluate how different methods scale as the number of queries to generate increases. We use ``Redset\_Cost\_Hard''  distribution with IMDB as the underlying database, fix the number of intervals to 10, and vary the number of queries from 50 to 500 and 5000. As shown in the first row of Figure~\ref{fig:scalability}, SQLBarber demonstrates significantly better scalability. Specifically, SQLBarber consistently reduces the distance to zero within tens of minutes, regardless of the number of queries. In contrast, baseline methods succeed only when the number of queries is small (e.g., 50), but fail to reduce the distance to zero when the number increases to 500 or 5000, even after hundreds of minutes. For baseline methods, query quality degrades as the number of queries increases, as indicated by larger final distances.

\noindent \textbf{Scaling with the number of intervals.} We evaluate how different methods scale as the number of intervals increases. We fix the number of queries to 1000, and vary the number of intervals from 5 to 10, 15, 20, and 25. As shown in the second row of Figure~\ref{fig:scalability}, SQLBarber again demonstrates significantly better scalability. In contrast, baseline methods are only effective when the number of intervals is as small as 5, and fail to converge as the number of intervals increases, even after hundreds of minutes. \textcolor{black}{We observe that for baseline methods, the distance does not always increase with more intervals. Since the total number of queries is fixed, adding more intervals reduces queries per interval, leading to smaller differences between the desired and actual counts, and thus lower distances.}

\subsection{Ablation Study} \label{sec:ablation}

\begin{figure}[btph]
    \centering
    \includegraphics[width=0.6\linewidth]{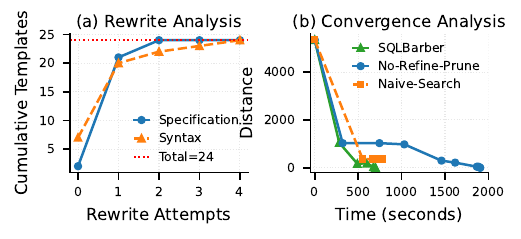}
    \caption{Ablation Study of SQLBarber (IMDB, Redset\_Cost)}
    \label{fig:abalation-study}
\end{figure}

\noindent \textbf{Effect of Customized SQL Template Generator.} \textcolor{black}{SQLBarber verifies query syntax using DBMS error messages and ensures consistency between specifications and templates through LLM-based self-correction (Section~\ref{sec:template-generator}, Step 5). This iterative rewrite process is automatically conducted, and to evaluate this process, we manually inspected whether the templates generated in different attempts satisfy their specifications. This manual inspection is only for evaluation purposes to produce Figure~\ref{fig:abalation-study}(a), which shows the cumulative number of templates that are correct with respect to the specification (blue) or syntax (orange) after each rewrite attempt.} Initially, only 2 templates meet the specification and 8 are syntactically correct. By the fourth attempt, all 24 templates are both correct and error-free. This demonstrates the effectiveness of our Algorithm~\ref{alg:template_check_rewrite} in guiding the LLM to self-correct through iterative hybrid feedback.

\noindent \textbf{Effect of Cost-Aware Query Generator.} Figure~\ref{fig:abalation-study}~(b) shows how the Wasserstein distance changes over time when using different versions of {SQLBarber}. The variant ``No-Refine-Prune'' disables Algorithm~\ref{alg:template_refinement}, so it does not adaptively refine SQL templates. The variant ``Naive-Search'' replaces our BO-based predicate search (Algorithm~\ref{algo:bo-predicate-search}) with a random search strategy. ``No-Refine-Prune'' takes approximately 3$\times$ longer to reduce the distance to zero because it cannot adapt SQL templates to complex distributions. ``Naive-Search'' fails to reduce the distance to zero because it cannot effectively select templates for different cost ranges or search for suitable predicate values. These results highlight the importance of both template refinement and BO-based predicate search.

\subsection{\textcolor{black}{Further Analysis}}~\label{subsec: further-anlaysis}

\noindent \textbf{\textcolor{black}{Cost Analysis.}}
\textcolor{black}{We report the number of tokens and the monetary cost incurred by SQLBarber and SQLStorm on the IMDB database in Table~\ref{tab:imdb-cost-sqlbarber-sqlstorm}. On average, SQLBarber incurs costs of 1.36 USD to generate thousands of queries. The cost varies across benchmarks because SQLBarber adaptively generates queries according to the task characteristics. In contrast, SQLStorm generates queries randomly without feedback, so its cost is independent of the benchmark. Although SQLStorm consumes about 80 times more tokens, its monetary cost is only 3.56 times higher because \texttt{gpt-4o-mini} is cheaper than \texttt{o3-mini} and benefits from a 50\% discount for batch processing provided by OpenAI. This difference also leads to a substantial increase in execution time: SQLStorm requires 1016 minutes, whereas SQLBarber takes approximately 18 minutes.}

\begin{table}[htbp]
\small
\centering
\caption{\textcolor{black}{Token Usage and Monetary Cost on IMDB}}
\label{tab:imdb-cost-sqlbarber-sqlstorm}
\begin{tabularx}{\columnwidth}{@{} l l >{\raggedleft\arraybackslash}X r @{}} 
\toprule
\textbf{System} & \textbf{Benchmark} & \textbf{Tokens (K)} & \textbf{Cost (USD)} \\
\midrule
SQLBarber & uniform               & 416    & 1.20 \\
SQLBarber & normal                & 514    & 1.50 \\
SQLBarber & Snowset\_Cost\_Medium & 387    & 1.22 \\
SQLBarber & Redset\_Cost\_Medium  & 513    & 1.49 \\
SQLBarber & Snowset\_Cost\_Hard   & 453    & 1.39 \\
SQLBarber & Redset\_Cost\_Hard    & 452    & 1.35 \\
\midrule
\textcolor{black}{SQLStorm}  & \textcolor{black}{Benchmark-Independent}     & \textcolor{black}{34\,735} & \textcolor{black}{4.84} \\
\bottomrule
\end{tabularx}
\end{table}

\begin{figure}[htbp]
    \centering
    \includegraphics[width=0.64\linewidth]{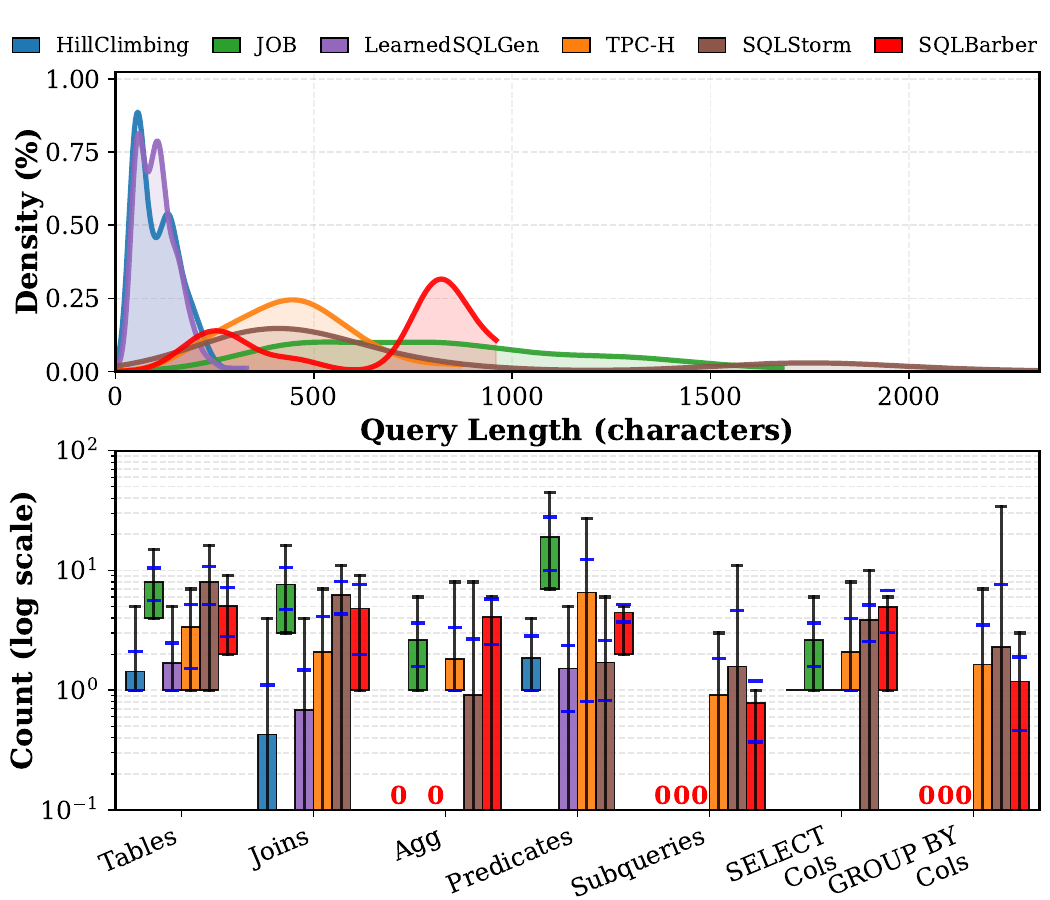}
    \caption{\textcolor{black}{Query Characteristics Comparison}}
    \label{fig:query-comparison}
\end{figure}

\noindent \textbf{\textcolor{black}{Comparison of Query Characteristics.}} \textcolor{black}{Figure~\ref{fig:query-comparison} compares the characteristics of queries from existing benchmarks and those generated by different systems. The top panel shows the distribution of query lengths, while the bottom panel presents the counts of structural metrics. In the bottom panel, colored bars represent mean values, black error bars indicate the minimum–maximum range, and blue bars denote the standard deviation. SQLBarber is not restricted to the shown distribution, as it can generate queries that satisfy user-specified characteristics. This capability is not available in other methods, whose generation processes are random and uncontrolled. Queries generated by all systems differ notably from those in existing benchmarks, which is desirable because such queries expose different challenges. HillClimbing and LearnedSQLGen produce the least diverse queries, while LLM-based approaches (SQLStorm and SQLBarber) achieve higher diversity, as indicated by the query length distributions and structural metric statistics. SQLBarber generates slightly less diverse queries than SQLStorm, as expected, because its generation process is guided by real-world distributions, whereas SQLStorm generates queries randomly.}

\begin{figure}[btph]
    \centering
    \includegraphics[width=0.64\linewidth]{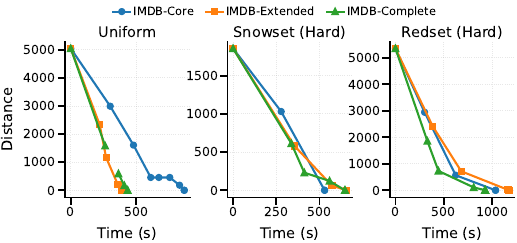}
    \caption{\textcolor{black}{Effect of Database Schema Complexity}}
    \label{fig:effect-schema-complexity}
\end{figure}

\noindent \textbf{\textcolor{black}{Effect of Schema Complexity.}} 
\textcolor{black}{As shown in Table~\ref{tab:synthetic-schemata}, we synthesize IMDB-derived databases of different schema complexity. This is achieved by progressively expanding a core schema and controlling attribute retention to vary structural complexity, with detailed procedures provided in the appendix due to space limitations. }
\textcolor{black}{Figure~\ref{fig:effect-schema-complexity} illustrates the effect of database schema complexity on SQLBarber when targeting three different cost distributions.
SQLBarber consistently reduces the distance to zero through its adaptive, feedback-driven query generation.
Although execution times vary across schemas, there is no clear correlation between schema complexity and runtime.
}


\begin{table}[t]
\centering
\small
\setlength{\tabcolsep}{2.2pt} 
\caption{\textcolor{black}{Schema Complexity of IMDB-Derived Databases}}
\label{tab:synthetic-schemata}
\begin{tabular}{lccccc}
\toprule
\textbf{Databases} & \textbf{\#Tables} & \textbf{\#Columns} &
\textbf{Avg Cols/Table} & \textbf{Join Range} \\
\midrule
IMDB-Core     & 9  & 27  & 3.0  & 1--7  \\
IMDB-Extended & 16 & 53  & 3.3  & 1--12 \\
IMDB-Complete & 21 & 81  & 3.9  & 1--16 \\
\bottomrule
\end{tabular}
\end{table}

\noindent \textbf{\textcolor{black}{More Target Cost Metrics.}}
\textcolor{black}{
SQLBarber can be extended to optimize for additional cost metrics. As an example, we consider a CPU usage–based cost function. The total CPU cost is defined as the sum of CPU costs over all nodes, and we reuse the node-specific CPU cost formulas from PostgreSQL (\texttt{costsize.c}). More details about this function are provided in the Appendix. As shown in Figure~\ref{fig:cpu-experiment}, we use SQLBarber to generate queries on the IMDB and TPC-H databases based on a real-world CPU usage distribution from Snowset~\cite{snowflake-nsdi20}, and compare the results against other cost metrics. SQLBarber reduces the distance to 0 within 30 minutes at a low monetary cost (\$1.2 on IMDB and \$1.7 on TPC-H), achieving comparable performance across different metrics. These results show that SQLBarber can effectively adapt to various cost functions through adaptive template generation and query instantiation.}

\begin{figure}[btph]
    \centering
    \includegraphics[width=0.64\linewidth]{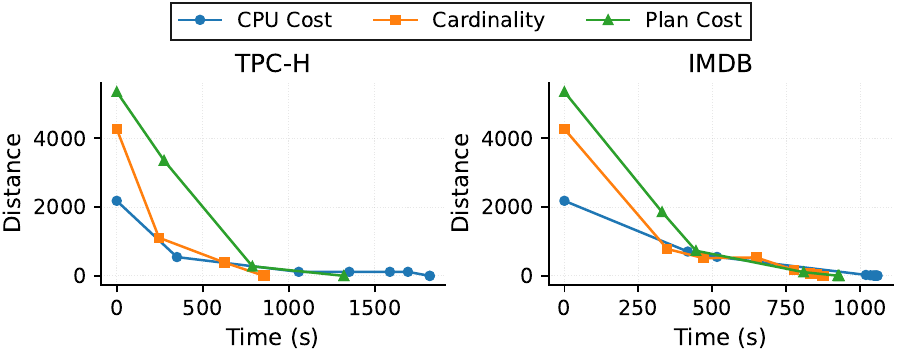}
    \caption{\textcolor{black}{SQLBarber Performance on Different Cost Metrics}}
    \label{fig:cpu-experiment}
\end{figure}



\begin{figure}[htbp]
    \centering
    \includegraphics[width=0.45\linewidth]{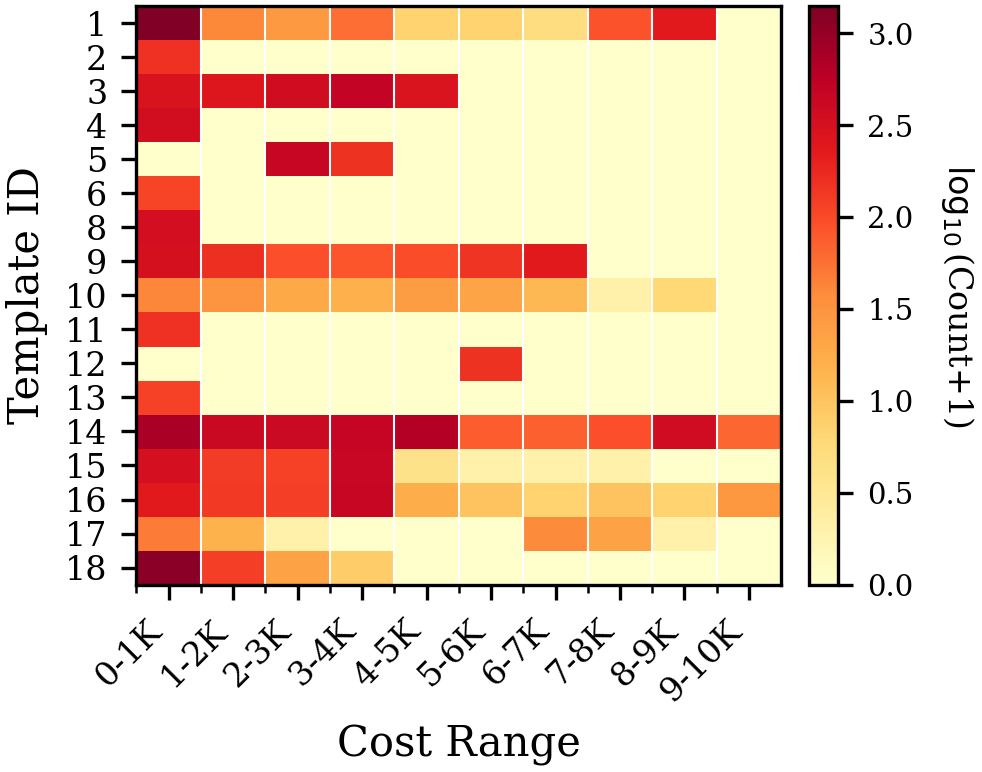}

    \caption{\textcolor{black}{Relation Between Templates and Cost Ranges}}
    \label{fig:template-cost-heatmap}
\end{figure}

\noindent \textbf{\textcolor{black}{Relation Between Template Specifications and Cost Ranges.}} \textcolor{black}{Figure~\ref{fig:template-cost-heatmap} shows a heatmap illustrating the cost ranges covered by each SQL template. We present 18 templates generated by SQLBarber from 18 template specifications in Redset~\cite{redset-vldb}. Some templates span all cost ranges (14 and 16), while others cover only narrow ranges (template 12 covers 5k–6k). Certain intervals are covered by many templates (0–1k by 16 templates), whereas others are covered by only a few (9–10k by templates 14 and 16), making those templates crucial. This figure shows that (1) SQLBarber produces diverse templates that cover different cost ranges with varying frequencies, and (2) a fixed template set cannot cover all ranges, motivating SQLBarber’s adaptive template refinement.}

\begin{figure}[btph]
    \centering
    \includegraphics[width=0.65\linewidth]{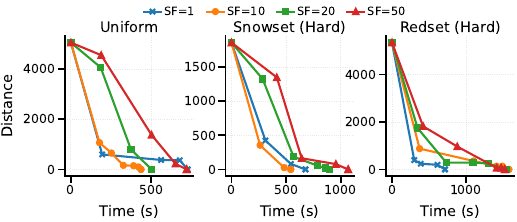}
    \caption{\textcolor{black}{Effect of Database Sizes (TPC-H, Query Plan Cost)}}
    \label{fig:effect-sizes}
\end{figure}

\noindent \textbf{\textcolor{black}{Effect of Database Types and Sizes.}} \textcolor{black}{We experimented on two databases, IMDB and TPC-H, to show that SQLBarber can adapt to different database types, which is verified by experiments in Figure~\ref{fig:distance-over-time}, Figure~\ref{fig:generation-time}, and Figure~\ref{fig:cpu-experiment}. For database sizes, as shown in Figure~\ref{fig:effect-sizes}, we vary the TPC-H SF from 1 to 50, including SF = 1, 10, 20, and 50. SQLBarber consistently reduces the distance to zero across all database sizes through its adaptive template generation and query instantiation. The optimization time, however, varies with database size because the database-independent seed templates exhibit different coverage over the target cost distribution. Specifically, after seed template generation and profiling, SQLBarber covers 17 and 14 out of 20 cost ranges for SF = 1 and SF = 10, but only 2 and 11 ranges for SF = 20 and SF = 50. For SF = 50, no queries appear in the small cost ranges (500–5000). This is expected, as generating low-cost queries becomes more difficult on larger databases. These uneven initial coverages lead to varying optimization difficulty in the refinement phase. Nevertheless, SQLBarber adaptively adjusts its templates based on feedback from the target database, enabling it to gradually fill all cost ranges and ultimately reduce the distance to zero. Moreover, we observe that there are always some queries falling within the lowest cost range (0–500), regardless of the data scale. To understand this, we analyze the cost ranges produced by each SQL template and find that, due to our random table selection strategy for diverse template generation, some templates access constant-size tables in TPC-H (e.g., region and nation). These queries are therefore inherently low-cost, independent of the database size. However, these queries only account for about 40\% of all queries in the smallest cost range. The remaining queries in this range either access tables with highly selective filters or combine constant tables with small tables under multiple predicate conditions. This diversity introduced by SQLBarber is beneficial, as it produces a wide variety of queries across different cost ranges rather than generating repetitive queries that only access constant-size tables.
}

\section{Conclusions and Outlook}\label{sec:conclusion}
We present SQLBarber, a system that uses Large Language Models (LLMs) to generate customized and realistic SQL workloads. Extensive experiments demonstrate its effectiveness, efficiency, and scalability, showing clear improvements over existing methods. We also perform an ablation study to analyze each algorithmic component and a cost study to assess its cost-effectiveness. Finally, we further analyze the characteristics of the generated queries, and evaluate the effects of schema complexity, database types, and database sizes on SQLBarber. We also study the use of CPU usage as a target cost metric and examine the relation between template specifications and cost ranges.

SQLBarber currently targets the OLAP scenario, but it can be extended to support OLTP workloads, including transaction generation and concurrency control. It can also be adapted to support cloud-native features such as multi-tenancy, query repetition, and temporal elasticity. Another interesting extension is enabling SQLBarber to generate semantic queries that extend SQL with semantic operators powered by LLMs or other machine learning models to process multi-modal data. Furthermore, our experiments show that the characteristics of the generated queries and their optimization difficulty are influenced by the types and sizes of the databases. This suggests that SQLBarber could be extended to jointly model and generate both databases and queries. Finally, a promising direction is to explore how to scale the system when using cost metrics that require the actual execution of SQL queries (e.g., execution time).

\bibliographystyle{ACM-Reference-Format}
\bibliography{sample-base}

\clearpage
\appendix

\section{\textcolor{black}{CPU Usage Estimation}}

\textcolor{black}{
SQLBarber can be extended to support other cost functions, such as CPU usage, memory consumption, and actual execution time. 
We demonstrate SQLBarber's flexibility using a cost function for estimated CPU usage. 
Although we focus on CPU costs in this example, SQLBarber is not restricted to this metric. 
To estimate CPU consumption, we adopt the \emph{CPU components} of PostgreSQL’s planner cost model and apply them to the plan that the optimizer produces via \texttt{EXPLAIN (FORMAT JSON)}. 
}

\textcolor{black}{
Let $P = \{n_1, n_2, \dots, n_k\}$ denote the plan nodes, each generating an estimated number of output tuples $r_i$. 
The total CPU cost is defined as the sum of per-node CPU costs:
\[
C_{\mathrm{CPU}}(P) = \sum_{n_i \in P} C_{\mathrm{CPU}}(n_i).
\]
}

\textcolor{black}{
We reuse the node-specific CPU formulas from PostgreSQL and exclude I/O-related terms (\texttt{costsize.c}). Typical cases include:}

\[
\textcolor{black}{
\begin{aligned}
C_{\mathrm{SeqScan}} &= (c_{\text{tuple}} + n_{\text{qual}}c_{\text{op}})\,r_{\text{rel}}, \\
C_{\mathrm{IndexScan}} &= (c_{\text{index}} + c_{\text{tuple}} + n_{\text{qual}}c_{\text{op}})\,r_{\text{cand}}, \\
C_{\mathrm{Sort}} &= 2c_{\text{op}}k\,r_{\text{in}}\log_2 r_{\text{in}}, \\
C_{\mathrm{HJ}} &= c_{\text{op}}\,n_{\text{hash}}\,(r_{\text{build}}+r_{\text{probe}}) + c_{\text{tuple}}\,(r_{\text{build}}+r_{\text{probe}}), \\
C_{\mathrm{MJ}} &\approx c_{\text{op}}\,n_{\text{merge}}\,(r_L+r_R) + c_{\text{tuple}}\,(r_L+r_R), \\
C_{\mathrm{NL}} &\approx c_{\text{tuple}}\,(r_L+r_R) + c_{\text{op}}\,n_{\text{qual}}\,(r_L+r_R), \\
C_{\mathrm{Agg}} &= c_{\text{tuple}}\,r_{\text{in}} + c_{\text{op}}\,n_{\text{group}}\,g.
\end{aligned}}
\]

\textcolor{black}{
Here $n_{\text{qual}}$ denotes the number of qualifiers or expressions evaluated in a node (\texttt{cost\_qual\_eval}); 
$k$ is the number of sort keys; $n_{\text{hash}}$ and $n_{\text{merge}}$ are the numbers of join predicates for hash and merge joins, respectively; 
$r_{\text{cand}}$ is the number of candidate tuples from an index; $r_{\text{rel}}$ is the base relation’s tuples; 
$r_{\text{in}}$ is the input size for sort or aggregate; $r_{\text{build}}$ and $r_{\text{probe}}$ are the build and probe relations in a hash join; 
$r_L$ and $r_R$ are inputs to merge or nested-loop joins; and $g$ is the number of output groups.}

\textcolor{black}{
PostgreSQL internally derives $n_{\text{qual}}$ by traversing the expression tree during planning. 
Because \texttt{EXPLAIN} only outputs textual predicates, we approximate $n_{\text{qual}}$ by counting logical and comparison operators 
(e.g., ``='', ``<'', ``>'', ``IN'', ``LIKE'') in each node’s filter or join condition. 
This approximation provides a consistent, relative measure of expression evaluation effort while keeping the original plan structure unchanged.}

\textcolor{black}{
We use the default GUC parameter values from PostgreSQL to compute all constants, as listed in Table~\ref{table:postgresql-cost-constants}. 
This CPU-only model preserves PostgreSQL’s cost formulation while excluding I/O terms, offering a consistent and interpretable cost estimate for query plans.}

\section{\textcolor{black}{Databases of Different Schema Complexity}}

\textcolor{black}{We synthesize IMDB-derived schemata of increasing complexity by (1) fixing a semantic backbone consisting of the titles–cast–people tables with their typed attributes, and then expanding the join graph with additional satellite tables (e.g., companies, aka-tables, links, and complete cast), and (2) controlling column width through a structured pruning policy. Specifically, we always retain key columns (primary and foreign keys), preserve a tiered subset of predicate columns (a few in Core, more in Extended, and all in Complete), and remove decorative columns (e.g., notes, codes, auxiliary identifiers) in lighter tiers. The data are generated by applying a project–prune transformation to the original IMDB/JOB load: each tier is materialized as a set of column projections and table removals over the base dataset, while maintaining foreign key integrity by keeping only rows whose referenced targets remain. This process produces comparable datasets where performance differences arise primarily from schema size and structure rather than data volume.}

\begin{table}[bpht]
\centering
\caption{\textcolor{black}{PostgreSQL Default Cost Constants}}\label{table:postgresql-cost-constants}
\begin{tabular}{lcl}
\hline
\textbf{Symbol} & \textbf{GUC Name} & \textbf{Default Value} \\
\hline
$c_{\text{seq\_page}}$ & \texttt{seq\_page\_cost} & 1.0 \\
$c_{\text{rand\_page}}$ & \texttt{random\_page\_cost} & 4.0 \\
$c_{\text{tuple}}$ & \texttt{cpu\_tuple\_cost} & 0.01 \\
$c_{\text{index}}$ & \texttt{cpu\_index\_tuple\_cost} & 0.005 \\
$c_{\text{op}}$ & \texttt{cpu\_operator\_cost} & 0.0025 \\
$c_{\text{parallel\_setup}}$ & \texttt{parallel\_setup\_cost} & 1000.0 \\
$c_{\text{parallel\_tuple}}$ & \texttt{parallel\_tuple\_cost} & 0.1 \\
\hline
\end{tabular}
\end{table}

\end{document}